\documentclass[11pt]{article}%
\usepackage{a4wide}
\usepackage{graphicx}
\usepackage{amsmath}
\usepackage{amsfonts}
\usepackage{amssymb}
\begin{document}

\title{Large amplitude spin waves in ultra-cold gases}
\author{J.N.\ Fuchs, D.M.\ Gangardt and F.\ Lalo\"{e}\\LKB, D\'{e}partement de Physique de l'ENS, \\24 rue Lhomond, 75005 Paris, France}
\date{\today}
\maketitle

\begin{abstract}
We discuss the theory of spin waves in non-degenerate ultra-cold gases, and
compare various methods which can be used to obtain appropriate kinetic
equations.\ We then study non-hydrodynamic situations, where the amplitude of
spin waves is sufficiently large to bring the system far from local
equilibrium. The full position and momentum dependence of the distribution
function must then be retained.

In the first part of the article, we compare two general methods which can be
used to derive a kinetic equation for a dilute gas of atoms (bosons or
fermions) with two internal states (treated as a pseudo-spin $1/2$). The
collisional methods are in the spirit of Boltzmann's original derivation of
his kinetic equation where, at each point of space, the effects of all sorts
of possible binary collisions are added. We discuss two different versions of
collisional methods, the Yvon-Snider approach and the $S$ matrix
approach.\ The second method uses the notion of mean field, which modifies the
drift term of the kinetic equation, in the line of the Landau theory of
transport in quantum liquids.\ For a dilute cold gas, it turns out that all
these derivations lead to the same drift terms in the transport equation, but
differ in the precise expression of the collision integral and in higher order
gradient terms.

In the second part of the article, the kinetic equation is applied to spin
waves (or internal conversion) in trapped ultra-cold gases. Numerical
simulations are used to illustrate the strongly non-hydrodynamic character of
the spin waves recently observed with trapped $^{87}$Rb atoms. The decay of
the phenomenon, which takes place when the system relaxes back towards
equilibrium, is also discussed, with a short comment on decoherence.

In two appendices we calculate the Wigner transform of the interaction term in
the $S$ matrix method, to first order in gradients; appendix I treats the case
of spin-independent interactions, appendix II that of spin-dependent interactions.

\end{abstract}

\section{Introduction}

Recent experiments \cite{Cornell, Cornell-2} have renewed the interest in spin
waves in dilute quantum gases, in conditions where they had not been observed
before.\ A first novelty is that these waves were neither purely nuclear nor
electronic, but involved two hyperfine atomic levels where the nuclear and
electronic spin are coupled.\ This difference is actually minor: the exchange
effets from which the waves originate are unaffected by the precise nature of
the levels - in other words one can, without any loss of generality,
assimilate any pair of atomic states to up and down states of a fictitious
spin.\ The second, more important, difference is that the waves were observed
with very large amplitudes, leading to an almost complete apparent segregation
of the atoms in each internal state, and involving situations where the gas is
very far from local equilibrium; we study such situations in the present article.

In degenerate liquids, the existence of spin waves has been known for many
years \cite{Silin,Leggett}. In dilute gases, which are non-degenerate, it took
about twenty more years \cite{Bashkin, LL2} to realize that they should also
sustain similar waves, for both bosonic and fermionic systems.\ This came as a
surprise to some since, before, these waves had been mostly associated with
the Landau formalism for Fermi liquids, and with the notion of
\textquotedblleft molecular field\textquotedblright\ resulting from exchange
and interactions with nearest neighbors.\ In a liquid, a test particle remains
trapped in a sort of local cavity and interacts constantly with several
neighbors, so that the notion of mean field emerges naturally from the
averaging over the effect of several neighbors.\ By contrast, in a gas, the
motion of a particle is often described as free flights along straight lines,
interrupted by short collisional processes, so that the physics involved is
clearly very different. It nevertheless turns out that, when averaged over all
possible collisions at each point of space, the final result of the exchange
interactions in a gas and a liquid are very similar (see, for example, Ref.
\cite{Miyake} for a general discussion). Soon after these theoretical
predictions, experimental observations demonstrated the existence of spin
waves in spin polarized Hydrogen gas \cite{Hpol}, $^{3}$He gas \cite{Hepol},
and dilute $^{3}$He-$^{4}$He solutions \cite{Gully}.

The physical difference between gases and liquids has its counterpart in the
different theoretical approaches used to derive kinetic
equations.\ Historically, the first derivation was that of Boltzmann, with his
Stosszahlansatz (or molecular chaos Ansatz): collisions are considered as
point processes (with no duration and no spatial extent) taking place between
particles which are completely uncorrelated before collision.\ As a
consequence, one can add at each point of space the effect of all possible
binary collisions between uncorrelated particles, as if one was adding the
effect of many \textquotedblleft beam to beam collision experiments" in atomic
physics.\ This provides the famous Boltzmann equation, with a relatively
complicated collision term on the right hand side, while the drift term on the
left hand side corresponds merely to a completely free flight of particles
between collisions. In the approach introduced later by Landau for the study
of degenerate Fermi liquids \cite{Kinetics}, quasi-particles are never free,
since their motion is constantly guided by a mean field created by the
neighbor particles.\ Therefore interactions modify the drift term while, on
the other hand, the collision term is generally treated phenomenologically, by
a simple relaxation approximation.\ Both point of view have their
advantages.\ Boltzmann's point of view does not require the use of a
pseudo-potential and therefore allows a more microscopic treatment of
collisions, for instance a more precise inclusion of lateral collisions
including their full angular end energy dependence.\ On the other hand, it
does not encompass quantum degenerate systems, does not introduce the powerful
notion of quasi-particles, and generally speaking remains limited to dilute
interacting gases.

The connection between the two points of view is provided by collisions in the
forward direction.\ In two body collision theory, it is well known that wave
interference in the forward direction is responsible for particle absorption
(optical theorem).\ In addition, more physics may be involved in this forward
interference effect: for spinless particles, retardation effects are
introduced (see for instance \cite{retardation}); for particles with spin, in
addition, identical spin rotation effects (ISRE) due to particle
indistinguishability \cite{LL1, LL2} also take place.\ In many-body transport
theory, both these effects have their equivalents.\ The cumulated retardation
effects during collisions give rise to an average force, or equivalently to a
scalar mean field (see for instance \cite{Mullin-Laloe} and references
therein); in the kinetic equation, this corresponds to terms which are
quadratic in the distribution function, as the collision term, but contain
gradients (spatial gradients of the density as well as momentum gradients of
the particle distribution); in other words, one gets zero sound Landau type
mean field terms which are naturally grouped with the drift term on the l.h.s.
of the kinetic equation.\ As for the identical spin rotation effects, they
average to a spin molecular field which is the equivalent of the molecular
field considered by Silin \cite{Silin} and Leggett \cite{Leggett}; the ISRE is
already present to zero order gradient expansion, as opposed to the scalar
retardation terms \cite{LL1, Meyerovich}, which explains why the dispersion
relation of spin waves is different from the usual sound wave dispersion.\ A
detailed study of forward scattering in a binary collision can therefore lead
to a microscopic understanding of the origin of the Landau mean field (lateral
scattering introduces collisional damping).\ From a practical point of view,
it remains true that the mean field treatment of the drift term, necessarily
associated with the use of a pseudo-potential, is often more compact and
elegant than a detailed study of the collisional properties in the forward
direction; it is nevertheless interesting to check precisely to what extent
they are equivalent in various situations.

Our purpose in this article is twofold. First, for particles with spins, the
equivalence in question has been verified in the literature only to zero order
in the gradient expansion, in other words only for the local terms appearing
in the kinetic equation. Here we wish to derive the kinetic equation to first
(non-local) order in the gradient expansion. Moreover, we will consider the
situation where the scattering may depend on the relative spin orientations
(even if this dependence was relatively weak in the experiments of the JILA
group \cite{Cornell, Cornell-2}).\ We will therefore examine in detail the
properties of collisional interference in the forward direction in this more
general case; the details of the calculations are given in two appendices
(gradient expansion of the interaction term within the $S$-matrix collision
Ansatz).\ Second, we also wish to study situation where the gas is far from
local equilibrium.\ In fact, most of the work on spin waves in the literature
deals with hydrodynamic situations where, at every point of space, the gas is
close to local equilibrium, so that a simple hydrodynamic expression of the
spin current can be used; of course, this reduces the number of variables, but
at the price of an approximation which is not necessarily justified.\ In fact,
in the experiments in question, the gas sometimes evolves very far from an
hydrodynamic regime, which is not so surprising since the spin waves have a
large amplitude and lead to an almost complete segregation of the two spin
species. For a study of spin waves in the collisionless regime (mean-free-path
large compared to characteristic lengths), but still close to equilibrium
(linear regime), see Ref. \cite{Bashkin2}.

\section{Kinetic equation}

We study an ensemble of identical atoms with two internal levels, obeying
either Bose or Fermi statistics.\ As mentioned in the introduction we can,
without any loss of generality, assimilate these two levels to two spin
levels, whatever their real physical origin is.\ For instance, hyperfine
atomic structure states, even in a situation of intermediate magnetic
decoupling, are possible; the only important thing is that the quantum states
describing the internal variables should be orthogonal.\ We now derive a
kinetic equation for a dilute gas of such atoms.

\subsection{Collisional methods; Yvon-Snider and Lhuillier-Lalo\"{e}
equations}

In classical statistical mechanics, Boltzmann's intuitive method
for deriving a kinetic equation is based on the study of
individual collisions; one adds the effect of all possible binary
collisions taking place at each point of
space on the single particle distribution in phase-space $f_{1}(\mathbf{r}%
,\mathbf{p})$.\ Another point of view starts from the infinite hierarchy of
BBGKY equations and closes its first equation, which relates $f_{1}$ to the
two particle distribution function $f_{2}$, by studying the evolution of
$f_{2}$ during a binary collision and expressing it as some functional of
$f_{1}$ (see for instance Ref. \cite{McLennan, Zubarev}).

The same lines can be followed in quantum mechanics, but $f_{1}(\mathbf{r}%
,\mathbf{p})$ has to be replaced by the reduced single-particle density
operator $\hat{\rho}_{1}$. Since we assume here that the particles have two
internal states, which we treat as spin states, the single-particle density
operator $\hat{\rho}_{1}$ acts in the product space of orbital and spin
variables of one particle. The operatorial kinetic equation has the form:
\begin{equation}
\frac{d\hat{\rho}_{1}}{dt}+\frac{1}{i\hbar}~\left[  \hat{\rho}_{1},\hat{H}%
_{1}\right]  _{-}\simeq\left.  \frac{d}{dt}\right\vert _{coll}\hat{\rho}_{1}
\label{2}%
\end{equation}
where:%
\begin{equation}
\hat{H}_{1}=\frac{\mathbf{p}^{2}}{2m}\hat{1}+\hat{V}^{ext} \label{3}%
\end{equation}
is the single-particle Hamiltonian, $\mathbf{p}$ the momentum of the particle,
$m$ its mass, $\hat{1}$ the unity operator in spin space and $\hat{V}^{ext}$
the operator describing the external forces acting on the particle, which may
be spin dependent.\ The r.h.s. of eq. (\ref{2}) takes into account the effects
of binary interactions between the particles; it contains the result of the
collisional approximation mentioned above, and will be discussed in more
detail below.\ The equation is only valid for a dilute gas ($n^{-1/3}\gg a$,
where $n$ is the number density of the gas and $a$ the scattering length) and
on time/length scales much greater than the duration of a collision/scattering length.

The next step is to introduce the Wigner transform \cite{Wigner, Wigner2}
$\hat{\rho}_{W}(\mathbf{r},\mathbf{p})$ of $\hat{\rho}_{1}$ with respect to
orbital variables:
\begin{equation}
\hat{\rho}_{W}(\mathbf{r},\mathbf{p})\equiv(2\pi\hbar)^{-3}\int\,d^{3}%
\mathbf{r}^{\prime}\,e^{i\mathbf{p}.\mathbf{r}^{\prime}/\hbar}~~\langle
\mathbf{r}-\frac{\mathbf{r}^{\prime}}{2}|\hat{\rho}_{1}|\mathbf{r}%
+\frac{\mathbf{r}^{\prime}}{2}\rangle\label{WignerTrans}%
\end{equation}
It is a classical function of position and momentum, but still an operator -
or a $2\times2$ matrix - in spin space. In the following, we will treat the
orbital degrees of freedom semi-classically, but not the spin degrees of
freedom.\ For the sake of simplicity, from now on we will drop the index $W$,
assuming that the dependence on $\mathbf{r}$ and $\mathbf{p}$ is sufficient to
signal a Wigner distribution. One then performs a Wigner transform of equation
(\ref{2}), which involves taking the Wigner transform of products of
operators.\ This is possible by using Groenewold's formula \cite{Groenewold,
Wigner2}, which provides the result as a infinite series of gradient
expansion: the first term is merely the product of the Wigner transforms of
the two operators, followed by product of gradients with respect to
$\mathbf{r}$ and $\mathbf{p}$, followed by higher order gradients, etc. The
Wigner transform of the l.h.s. of equation (\ref{2}) then gives:
\begin{align}
\partial_{t}\hat{\rho}(\mathbf{r},\mathbf{p})  &  +\frac{\mathbf{p}}{m}%
\cdot\nabla_{\mathbf{r}}\hat{\rho}(\mathbf{r},\mathbf{p})+\frac{1}{i\hbar
}[\hat{\rho}(\mathbf{r},\mathbf{p}),\widehat{V}^{ext}(\mathbf{r}%
)]_{-}\nonumber\\
&  -\frac{1}{2}[\nabla_{\mathbf{p}}\hat{\rho}(\mathbf{r},\mathbf{p}%
),\cdot\nabla_{\mathbf{r}}\widehat{V}^{ext}(\mathbf{r})]_{+}+...
\label{TWPauli}%
\end{align}
The first two terms are exact, whereas the next two are only the
first orders terms in a gradient expansion involving the external
potential. The commutator is obtained to zero order of the
gradient expansion while the anticommutator occurs only to first
order, followed by higher order terms symbolized by the dots. The
anticommutator corresponds to the classical force term (describing
the effect of a trap, for example), which we wish to retain in our
calculations; as a consequence, we have to include up to first
order gradient expansion in the Wigner transform of the r.h.s. of
equation (\ref{2}) .

The non-trivial part in the derivation of a kinetic equation
begins when one gives a precise expression to the formal r.h.s. of
equation (\ref{2}). This is the subject of many studies in the
literature and even the subject of books; here, we limit ourselves
to a simplified discussion of two different approaches which are
appropriate for the study of spin waves. We consider only
collisions taking place in elastic channels, where the number of
atoms in each internal spin state is conserved, and begin with
some simple considerations on elementary collision theory.\ For
each channel, at very low temperatures, collisions occur in the
extreme quantum regime where the typical $s$-wave scattering
length $a$ is small compared to the de Broglie thermal wavelength
$\lambda_{T}$. In this case, collisions are well described by the
isotropic $s$-wave, and the $T$ matrix can be approximated by:
\begin{equation}
T_{k}(\hat{k},\hat{k}^{\prime})=-\frac{\hbar^{2}}{2\pi^{2}m}f_{k}%
(\theta)=\frac{\hbar^{2}}{2\pi^{2}m}(a-ika^{2}+..) \label{1}%
\end{equation}
(terms of order $a^{2}$ should be retained in order to satisfy explicitly the
unitarity of the $S$-matrix in the optical theorem); in this equation, $m$ is
the mass of the particles, $f_{k}(\theta)$ the scattering amplitude which, for
$s$-wave scattering, is independent of the polar angle $\theta$, and $k$ the
collision wave vector.

A first obvious remark is that the intensity of the spherical scattered wave
is proportional to the square of $a$, while interference effects in the
forward direction between the incoming plane wave and the spherical scattered
wave are proportional to $a$ itself.\ Generally speaking, the relative
magnitude of forward and lateral scattering effects is of order $\lambda
_{T}/a$, which is large at low energies. This is why, here, we concentrate on
forward scattering effects and their relation to mean field corrections to the
drift terms. Another remark is that, because waves interfere only if they are
in phase, phases and $i$ factors are important here.\ It turns out that, in
quantum mechanics as well as in optics \cite{Feynman, Newton}, the summation
over many scatterers in the forward direction introduces an additional $i$
factor (which is actually the origin of the $i$ factor in the optical
theorem).\ Hence, in every scattering channel, equation (\ref{1}) shows that
no first order $a$ interference effect occurs in the forward direction, the
first contribution arising from the second term in $ika^{2}$; to first order
in $a$, only the phase of the forward wave is changed, not its
intensity.\ But, if several scattering channels are open, the situation is
different, because these different phase shifts in the forward scattered waves
can introduce spin rotation.\ Actually, this is precisely the origin of the
first order identical spin rotation effect in the forward direction \cite{LL1,
LL2}.

\subsubsection{Spin-independent interactions}

We now briefly summarize two approaches that have been developed in the
context of particles with spin: the Yvon-Snider approximation, which
emphasizes more the relation to the BBGKY hierarchy, and closes the infinite
set of equations with an approximation suitable for a binary collision; the
Lhuillier-Lalo\"{e} (or $S$ matrix) approach, which is closer to the initial
approach of Boltzmann and makes use of the $S$ collision matrix. We begin with
the simplest case, when the interactions are independent of spin.

\paragraph{Yvon-Snider}

An interesting method is that originally proposed by Yvon \cite{Yvon}, then
independently by Snider \cite{Snider}; it provides a result which, in the
literature, is often called the Waldmann-Snider equation;\ the main idea is to
express the collision term as the trace over a collision partner (particle
$2$) of a commutator containing the binary interaction potential $V_{12}$ and
a product of single particle operators modified by the unitary transformation
associated with the M{\o }ller collision operator $\Omega\equiv\Omega^{(+)}$:%
\begin{equation}
\left.  \frac{d}{dt}\right\vert _{coll}\hat{\rho}_{1}(1)=~(i\hbar)^{-1}%
~Tr_{2}\left\{  \left[  V_{12}~,~\Omega~\hat{\rho}_{1}(1)\hat{\rho}%
_{1}(2)~\Omega^{\dagger^{{}}}\right]  _{-}\right\}  \label{Yvon}%
\end{equation}
This equation is valid for dilute gases of distinguishable particles; for
non-degenerate gases ($n\lambda_{T}^{3}\ll1$, where $\lambda_{T}$ is the
thermal wavelength of the particles) of bosons or fermions, it generalizes
into:%
\begin{equation}
\left.  \frac{d}{dt}\right\vert _{coll}\hat{\rho}_{1}(1)=~(i\hbar)^{-1}%
~Tr_{2}\left\{  \frac{\hat{1}+\epsilon\hat{P}_{ex.}}{\sqrt{2}}\left[
V_{12}~,~\Omega~\hat{\rho}_{1}(1)\hat{\rho}_{1}(2)~\Omega^{\dagger^{{}}%
}\right]  _{-}\frac{\hat{1}+\epsilon\hat{P}_{ex.}}{\sqrt{2}}\right\}
\label{Yvon2}%
\end{equation}
where $\hat{P}_{ex.}$ is the exchange operator between particles $1$ and $2$,
and where:
\begin{equation}
\epsilon=1~\text{for bosons, }\epsilon=-1~\text{for fermions} \label{epsi}%
\end{equation}
Wigner transforms can then be applied to these expressions to complete the
derivation of the closed kinetic equation for $\hat{\rho}(\mathbf{r}%
,\mathbf{p})$.\ For instance, expanding (\ref{Yvon}) in a Groenewold gradient
expansion of the products provides, to zero order in gradients, the
straightforward Boltzmann expression for the collision term:
\begin{align}
\left.  \frac{d}{dt}\right\vert _{coll}\hat{\rho}(\mathbf{r}_{1}%
,\mathbf{p}_{1})  &  =-\int d^{3}q\frac{q}{m}~~\bigg[\sigma_{T}(k)~\hat{\rho
}(\mathbf{r}_{1},\mathbf{p}_{1})~f(\mathbf{r}_{1},\mathbf{p}_{2})\nonumber\\
&  \left.  -\int d^{2}\hat{k}^{\prime}~\sigma_{k}(\theta)~\hat{\rho
}(\mathbf{r}_{1},\mathbf{p}_{1}^{\prime})~f(\mathbf{r}_{1},\mathbf{p}%
_{2}^{\prime})\right]  \label{Boltzmann}%
\end{align}
where $~f(\mathbf{r},\mathbf{p})$ is the spin trace of $\hat{\rho}%
(\mathbf{r},\mathbf{p})$:%
\begin{equation}
f(\mathbf{r},\mathbf{p})=\text{Tr}_{S}\left\{  \hat{\rho}(\mathbf{r}%
,\mathbf{p})\right\}  \label{def-f}%
\end{equation}
and where $\hat{k}$ is the unit vector in the direction of $\mathbf{k}$ and
$\theta$ the angle between $\mathbf{k}$ and $\mathbf{k}^{^{\prime}}$. Here we
use the usual notation in Boltzmann theory:%
\begin{equation}%
\begin{array}
[c]{cc}%
\mathbf{q}=2\hbar\mathbf{k} & \mathbf{p}_{2}=\mathbf{p}_{1}-\mathbf{q}\\
\mathbf{p}_{1}^{\prime}=\mathbf{p}-(\mathbf{q}/2)+\hbar k\hat{k}^{\prime} &
\mathbf{p}_{2}^{\prime}=\mathbf{p}-(\mathbf{q}/2)-\hbar k\hat{k}^{\prime}%
\end{array}
\label{impulsions}%
\end{equation}
as well as the definition of the differential and total cross sections:
\begin{equation}%
\begin{array}
[c]{cc}%
\sigma_{k}(\theta)=(4\pi^{4}m^{2}/\hbar^{4})~\left\vert T(\mathbf{k}%
,\mathbf{k}^{\prime})\right\vert ^{2}; & \sigma_{T}(k)=\int d^{2}\hat
{k}^{\prime}~\sigma_{k}(\theta)
\end{array}
\label{sigmas}%
\end{equation}
Higher order terms will introduce scalar mean field terms, which we do not
write here explicitly for concision; they can be found in the appendix of
ref.\ \cite{Mullin-Laloe}.

Now, if we introduce spin and statistics, we have to use (\ref{Yvon2}) instead
of (\ref{Yvon}).\ To zero order in gradients, this introduces the additional
exchange terms to the right hand side of (\ref{Boltzmann}):%
\begin{equation}%
\begin{array}
[c]{r}%
\displaystyle-\frac{\epsilon}{2}\int d^{3}q~\frac{q}{m}~\Big\{i\tau
_{fwd.}^{ex.}~(k)~~\left[  \hat{\rho}(\mathbf{r}_{1},\mathbf{p}_{1}%
),\hat{\rho}(\mathbf{r}_{1},\mathbf{p}_{2})\right]  _{-}\\
-\int d^{2}\hat{k}^{\prime}~i\tau_{k}^{ex.}(\theta)~\left[  \hat{\rho
}(\mathbf{r}_{1},\mathbf{p}_{1}^{\prime}),~\hat{\rho}(\mathbf{r}%
_{1},\mathbf{p}_{2}^{\prime})\right]  _{-}+\\
\left.  +\int d^{2}\hat{k}^{\prime}~\sigma_{k}^{ex.}(\theta)\left(  \left[
\hat{\rho}(\mathbf{r}_{1},\mathbf{p}_{1}),\hat{\rho}(\mathbf{r}_{1}%
,\mathbf{p}_{2})\right]  _{+}-\left[  \hat{\rho}(\mathbf{r}_{1},\mathbf{p}%
_{1}^{\prime}),\hat{\rho}(\mathbf{r}_{1},\mathbf{p}_{2}^{\prime})\right]
_{+}\right)  \right\}
\end{array}
\label{exch}%
\end{equation}
where the \textquotedblleft generalized cross sections" are defined by:%
\begin{equation}
\tau_{fwd.}^{ex.}(k)=(-8\pi^{3}m/\hbar^{2}k)~\operatorname{Re}\left\{
T(-\mathbf{k},\mathbf{k})\right\}  \label{fwd}%
\end{equation}
and:%
\begin{equation}
\sigma_{k}^{ex.}(\theta)-i\tau_{k}^{ex.}(\theta)~=(4\pi^{4}m^{2}/\hbar
^{4})~T(-\mathbf{k},\mathbf{k}^{^{\prime}})~T^{\ast}(\mathbf{k},\mathbf{k}%
^{^{\prime}}) \label{lat}%
\end{equation}
In (\ref{exch}), the two terms in the first two lines correspond respectively
to the ISRE in the forward and lateral direction; the two terms in the third
line correspond to exchange effects changing the values of the total and
lateral cross sections, as discussed in more detail in \cite{LL1}. For
simplicity, we do not write the first order gradient terms, but more details
on the limit of validity of the Yvon-Snider equation and of the various terms
which it contains can be found in \cite{TS}, \cite{Meyerovich} or
\cite{Mullin-Laloe} (in particular the Appendix of this last reference).
Inserting the low energy limit (\ref{1}) is trivial; we write the explicit
results in the next section.

\paragraph{Lhuillier-Lalo\"{e}}

Another point of view, more directly in the spirit of the initial Ansatz of
Boltzmann, was employed by Lhuillier and Lalo\"{e} (LL) \cite{LL1}; it is
based on the use of the $S$ collision matrix.\ When collisions are treated
only as \textquotedblleft closed processes\textquotedblright\ (one ignores
particles \textquotedblleft in the middle of a collision\textquotedblright),
it is indeed possible to use the $S$ matrix to relate exactly the single
particle operators after and before collision.\ One gets the following
expression for the single particle density operator after collision $\hat
{\rho}_{1}^{\prime}$ as a function of the same operator $\hat{\rho}_{1}$
before collision:
\begin{equation}
\hat{\rho}_{1}^{\prime}(1)=Tr_{2}\left\{  \frac{\hat{1}+\epsilon\hat{P}_{ex.}%
}{\sqrt{2}}~\hat{S}~\hat{\rho}_{1}(1)~\hat{\rho}_{1}(2)~\hat{S}^{\dagger
}~\frac{\hat{1}+\epsilon\hat{P}_{ex.}}{\sqrt{2}}\right\}  \label{rho-prime}%
\end{equation}
(the trace acts over orbital and spin variables of the collision partner $2$)
where the $S$ matrix is related to the collision $T$ matrix by:%
\begin{equation}
\langle1:\mathbf{k}_{f};2:-\mathbf{k}_{f}~|~\hat{S}~|~1:\mathbf{k}%
_{i};2:-\mathbf{k}_{i}\rangle=\delta(\mathbf{k}_{f}-\mathbf{k}_{i})-i\frac{\pi
m}{\hbar^{2}k_{i}}~\delta(k_{f}-k_{i})~T_{k_{i}} \label{S}%
\end{equation}
($\hbar k$ is the relative momentum of the two particles). One then
approximates the rate of change of $\hat{\rho}_{1}$ by the variation
$(\hat{\rho}_{1}^{\prime}-\hat{\rho}_{1})/\Delta t$ during a short time
interval \footnote{We will see that $\Delta t$ is a time which emerges
naturally from the calculation through the occurrence of square of delta
functions of the energy (see Appendix I).} $\Delta t$, which is however much
longer than the duration of a collision, to obtain:
\begin{equation}
\left.  \frac{d}{dt}\right\vert _{coll}\hat{\rho}_{1}(1)\simeq\frac{1}{\Delta
t}Tr_{2}\left\{  \frac{\hat{1}+\epsilon\hat{P}_{ex.}}{\sqrt{2}}\left[  \hat
{S}~\hat{\rho}_{1}(1)~\hat{\rho}_{1}(2)~\hat{S}^{\dagger}-\hat{\rho}%
_{1}(1)~\hat{\rho}_{1}(2)\right]  \frac{\hat{1}+\epsilon\hat{P}_{ex.}}%
{\sqrt{2}}\right\}  \label{rhocoll}%
\end{equation}
One should note that the $S$ matrix Ansatz of equation (\ref{rhocoll}) is only
valid for a non-degenerate gas where:
\begin{equation}
Tr_{2}\left\{  \frac{\hat{1}+\epsilon\hat{P}_{ex.}}{\sqrt{2}}\hat{\rho}%
_{1}(1)~\hat{\rho}_{1}(2)\frac{\hat{1}+\epsilon\hat{P}_{ex.}}{\sqrt{2}%
}\right\}  =\hat{\rho}_{1}(1)+\epsilon\hat{\rho}_{1}(1)^{2}\simeq\hat{\rho
}_{1}(1) \label{Trrho}%
\end{equation}

As above, the kinetic equation is obtained from equation (\ref{rhocoll}) by
performing a Wigner transform followed by a gradient expansion. In Appendix~I
we first calculate the zero order gradient expansion, which is given by
equation (\ref{ILL}) ; it is easy to see that it coincides exactly with the
sum of (\ref{Boltzmann}) and (\ref{exch}), which shows that the Yvon-Snider
and $S$ matrix approximations are exactly equivalent to this order.\ The first
order gradient terms are also given in the same Appendix; here, we only give
their low energy limit by using equation (\ref{1}), which provides the
following kinetic equation:
\begin{align}
&  \partial_{t}\hat{\rho}(\mathbf{r}_{1},\mathbf{p}_{1})+\frac{\mathbf{p}_{1}%
}{m}\cdot\nabla_{\mathbf{r}}\hat{\rho}(\mathbf{r}_{1},\mathbf{p}_{1})+\frac
{1}{i\hbar}\left[  \hat{\rho}(\mathbf{r}_{1},\mathbf{p}_{1}),\widehat{V}%
^{ext}(\mathbf{r}_{1})+\epsilon\,g\hat{n}(\mathbf{r}_{1})\right]
_{-}\nonumber\\
&  -\frac{1}{2}\left[  \nabla_{\mathbf{p}}\hat{\rho}(\mathbf{r}_{1}%
,\mathbf{p}_{1}),\cdot\nabla_{\mathbf{r}}\big(\widehat{V}^{ext}(\mathbf{r}%
_{1})+gn(\mathbf{r}_{1})\hat{1}+\epsilon\,g\hat{n}(\mathbf{r}_{1}%
)\big )\right]  _{+}=I_{coll}[\hat{\rho}] \label{kinetic0}%
\end{align}
where $g\equiv4\pi\hbar^{2}a/m$ and $\hat{n}(\mathbf{r}_{1})$ is the local
density operator integrated over velocities:%
\begin{equation}
\hat{n}(\mathbf{r})\equiv\int d^{3}p~\hat{\rho}(\mathbf{r},\mathbf{p}) \label{n}%
\end{equation}
In this equation, as usual, forward scattering terms linear in the coupling
constant $g\propto a$ have been included in the l.h.s. The local term of the
collision integral, which regroups terms proportional to $a^{2}$, is given
by:
\begin{align}
I_{coll}[\hat{\rho}]  &  =-\int d^{3}q\,\frac{q}{m}\int d^{2}\hat{k}^{\prime
}\,a^{2}\Big(\hat{\rho}(\mathbf{r}_{1},\mathbf{p}_{1})f(\mathbf{r}%
_{1},\mathbf{p}_{2})-\hat{\rho}(\mathbf{r}_{1},\mathbf{p}_{1}^{\prime
})f(\mathbf{r}_{1},\mathbf{p}_{2}^{\prime})\nonumber\\
&  +\frac{\epsilon}{2}\left[  \hat{\rho}(\mathbf{r}_{1},\mathbf{p}_{1}%
),\hat{\rho}(\mathbf{r}_{1},\mathbf{p}_{2})\right]  _{+}-\frac{\epsilon}%
{2}\left[  \hat{\rho}(\mathbf{r}_{1},\mathbf{p}_{1}^{\prime}),\hat{\rho
}(\mathbf{r}_{1},\mathbf{p}_{2}^{\prime})\right]  _{+}\Big) \label{Icoll}%
\end{align}
where $\mathbf{p}_{2}$, $\mathbf{p}_{1}^{\prime}$ and
$\mathbf{p}_{2}^{\prime }$ are defined in (\ref{impulsions}). In
addition to the anticommutator in the l.h.s. of (\ref{kinetic0}),
obtained to first order in the gradient expansion, we recover
exactly the low-energy limit of the LL equation \cite{LL1} with,
in the l.h.s., the ISRE in the forward direction (the ISRE in the
lateral directions disappears in the low energy limit); the second
line of (\ref{Icoll}) provides the statistical terms introduced by
particle indistinguishability into the total and differential
collision cross section.

\subsubsection{Spin-dependent interactions}

We now study the gas with spin-dependent interactions and, for a moment,
consider the particles as distinguishable. In the limit of slow elastic
collisions, we need only consider four $s$-wave scattering lengths: $a_{11}$,
$a_{22}$, $a_{12}^{(d)}$ and $a_{12}^{(t)}$. The two first describe collisions
processes between particles that are in the same internal state; when they are
in othogonal internal states, two different processes occur for
distinguishable particles: a direct collision process without energy transfer
described by $a_{12}^{(d)}$, and a transfer collision process described by
$a_{12}^{(t)}$ - similar processes occur in the theory of spin exchange
collisions \cite{PL}. The $T$ matrix is now a $4\times4$ matrix in spin space;
to lowest order in the scattering lengths, its matrix elements with
distinguishable particles are given by:
\begin{equation}
\langle~1:\alpha~;2:\beta|~\hat{T}_{k}~|~1:\gamma~;2:\delta\rangle=\frac
{\hbar^{2}}{2\pi^{2}m}\left(
\begin{array}
[c]{cccc}%
\widetilde{a}_{11}(k) & 0 & 0 & 0\\
0 & \widetilde{a}_{22}(k) & 0 & 0\\
0 & 0 & \widetilde{a}_{12}^{(d)}(k) & \widetilde{a}_{12}^{(t)}(k)\\
0 & 0 & \widetilde{a}_{12}^{(t)}(k) & \widetilde{a}_{12}^{(d)}(k)
\end{array}
\right)  \label{Tspin}%
\end{equation}
where $\alpha$, $\beta$, $\gamma$ and $\delta$ are the internal state indexes
taking values $\{1,2\}$ and where:%
\begin{equation}
\widetilde{a}_{\alpha\beta}(k)=a_{\alpha\beta}\left[  1-ika_{\alpha\beta
}\right]  \label{atilde}%
\end{equation}
Equation (\ref{Tspin}) replaces (\ref{1}) when the interactions depend on spin.

\paragraph{Kinetic equation}

The derivation of the kinetic equation then proceeds along the same lines as
in the previous section: we use (\ref{rho-prime}) again to symmetrize the
density operator and take quantum statistics into account, but now insert
(\ref{Tspin}) instead of (\ref{S}) in it, calculate the partial trace, and
obtain an interaction term for the single particle density operator; we then
Wigner transform the result and use a gradient expansion limited to first
order. This calculation is similar to that of Appendix I for spin-independent
interactions, but of course more complicated; it provides more general
expressions for the mean field terms and the collision integrals. For
conciseness, we limit ourselves to the mean field terms, which will actually
be sufficient for our calculations below.\ In this particular case, the terms
can actually be obtained more conveniently by an operatorial method, different
from that of Appendix I, as shown in Appendix II.\ We only reproduce the final
kinetic equation:
\begin{align}
\partial_{t}\hat{\rho}(\mathbf{r},\mathbf{p})  &  +\frac{\mathbf{p}}{m}%
\cdot\nabla_{\mathbf{r}}\hat{\rho}(\mathbf{r},\mathbf{p})+\frac{1}{i\hbar
}[\hat{\rho}(\mathbf{r},\mathbf{p}),\widehat{U}(\mathbf{r})]_{-}\nonumber\\
&  -\frac{1}{2}[\nabla_{\mathbf{p}}\hat{\rho}(\mathbf{r},\mathbf{p}%
),\cdot\nabla_{\mathbf{r}}\widehat{U}(\mathbf{r})]_{+}=I_{coll}[\hat{\rho}]
\label{kinetic}%
\end{align}
where the effective single-particle potential $\widehat{U}$ is:
\begin{equation}
\widehat{U}=\widehat{V}^{ext}+\left(
\begin{array}
[c]{cc}%
(1+\epsilon)g_{22}n_{2}+g_{12}n_{1} & \epsilon g_{12}n_{21}\\
\epsilon g_{12}n_{12} & (1+\epsilon)g_{11}n_{1}+g_{12}n_{2}%
\end{array}
\right)  \label{U}%
\end{equation}
with $g_{\alpha\beta}\equiv4\pi\hbar^{2}a_{\alpha\beta}/m$, where
$a_{12}\equiv a_{12}^{(d)}+\epsilon a_{12}^{(t)}$; it turns out that only this
combination of the direct and transfer coupling constants enters the
calculation (for more details, see Appendix II).

This kinetic equation resembles the Landau-Silin equation \cite{Silin,LPL9}
for a normal Fermi liquid in a magnetic field. The second term in the l.h.s.
of (\ref{kinetic}) is the usual free drift term; the anticommutator is a force
term including both the effect of an external potential and of the mean field;
the commutator is a spin precession term containing effective magnetic fields
from different origins (differential Zeeman effect, spin mean field, etc.).

\paragraph{Evolution of the density and spin density in phase space}

The structure of the distribution function, which is still a matrix in the
space of internal variables, is more explicit if we decompose $\hat{\rho}$ and
$\hat{U}$ in the basis formed by Pauli matrices $\hat{\boldsymbol{\sigma}}$
and the unit $2\times2$ matrix:
\begin{align}
\hat{\rho}(\mathbf{r},\mathbf{p},t) &  =\frac{1}{2}\left(  f(\mathbf{r}%
,\mathbf{p},t)\hat{1}+\mathbf{M}(\mathbf{r},\mathbf{p},t)\cdot\hat
{\boldsymbol{\sigma}}\right)  \nonumber\\
\hat{U}(\mathbf{r},t) &  =U_{0}(\mathbf{r},t)\hat{1}+\mathbf{U}(\mathbf{r}%
,t)\cdot\hat{\boldsymbol{\sigma}}%
\end{align}
When written in terms of the phase-space density $f$ and spin density
$\mathbf{M}$, equation (\ref{kinetic}) becomes:
\begin{align}
\partial_{t}f+\frac{\mathbf{p}}{m}\cdot\nabla_{\mathbf{r}}f-\nabla
_{\mathbf{r}}U_{0}\cdot\nabla_{\mathbf{p}}f-\nabla_{r}\mathbf{U}\cdot
\nabla_{\mathbf{p}}\mathbf{M} &  =I_{coll}^{(f)}[f,\mathbf{M}]\nonumber\\
\partial_{t}\mathbf{M}+\frac{\mathbf{p}}{m}\cdot\nabla_{\mathbf{r}}%
\mathbf{M}-\nabla_{\mathbf{r}}U_{0}\cdot\nabla_{\mathbf{p}}\mathbf{M}%
-\nabla_{\mathbf{r}}\mathbf{U}\cdot\nabla_{\mathbf{p}}f-\frac{2\mathbf{U}%
}{\hbar}\times\mathbf{M} &  =I_{coll}^{(M)}[f,\mathbf{M}]\label{kinetic1}%
\end{align}
In these equations, $U_{0}$ plays the role of the overall trapping potential
and $\mathbf{U}$ is the local effective magnetic field around which the spins
rotate. The effective trapping potential $U_{0}$ is given by:
\begin{align}
U_{0}(\mathbf{r},t) &  =\frac{V_{2}^{ext}+V_{1}^{ext}}{2}+\left[
\frac{1+\epsilon}{2}(g_{22}+g_{11})+g_{12}\right]  \frac{n}{2}\nonumber\\
&  +\frac{1+\epsilon}{2}(g_{22}-g_{11})\frac{m_{\parallel}}{2}\label{trappot}%
\end{align}
and the effective magnetic field is:%
\begin{equation}%
\begin{array}
[c]{cl}%
\mathbf{U}(\mathbf{r},t)= & \displaystyle\frac{\hbar\Omega(x,t)}{2}%
\mathbf{e}_{\parallel}+\epsilon\frac{g_{12}\mathbf{m}(x,t)}{2}\\
\hbar\Omega(\mathbf{r},t)= & \displaystyle V_{2}^{ext}-V_{1}^{ext}%
+\frac{1+\epsilon}{2}\left[  (g_{22}-g_{11})n+(g_{22}+g_{11}-2g_{12}%
)m_{\parallel}\right]
\end{array}
\label{magn}%
\end{equation}
where the local density and spin polarization density are defined as:
\begin{align}
n=n_{2}+n_{1}\, &  ;\,m_{\parallel}=n_{2}-n_{1}\nonumber\\
m_{\perp,1}=n_{21}+n_{12}\, &  ;\,m_{\perp,2}=i(n_{21}-n_{12})
\end{align}
The basis in spin space is denoted by $\{e_{\bot,1}$; $e_{\bot,2}$;
$e_{\parallel}\}$. The two first vectors define the transverse plane, the last
defines the longitudinal direction.

The collision integral in (\ref{kinetic}), i.e. the equivalent of
eq. (\ref{Icoll}) when the interactions are spin-dependent, is not
written explicitly; to simplify the calculations, and because we
will mostly study experimental situations where lateral collisions
do not play a dominant role, we limit ourselves to a simple
relaxation time approximation:
\begin{equation}
I_{coll}^{(f)}\simeq-\frac{f(\mathbf{r},\mathbf{p},t)-f^{eq}(\mathbf{r}%
,\mathbf{p},t)}{\tau}\text{; }I_{coll}^{(M)}\simeq-\frac{\mathbf{M}%
(\mathbf{r},\mathbf{p},t)-\mathbf{M}^{eq}(\mathbf{r},\mathbf{p},t)}{\tau
}\label{tauapprox2}%
\end{equation}
where $f^{eq}$ (resp. $\mathbf{M}^{eq}$) is the local equilibrium (spin)
density in phase-space, and $\tau$ is a relaxation time. Actually, one could
allow for different relaxation times for $f$, $\mathbf{M}_{\parallel}$ and
$\mathbf{M}_{\perp}$. Nevertheless, in the case of a non-degenerate gas with
spin-independent interactions, it is known (Ref. \cite{tau} and references
therein) that these relaxation times are equal; below we consider a
non-degenerate gas of $^{87}$Rb, for which the three scattering lengths
$a_{11}$, $a_{22}$ and $a_{12}$ are very close \cite{Cornell}, so that in
practice we will we only need a single relaxation time $\tau$ given
approximately by:
\begin{equation}
\tau\simeq\frac{1}{4\pi a_{12}^{2}~n}\sqrt{\frac{m}{k_{B}T}}\label{tau}%
\end{equation}
Finally, the kinetic equations we will use below to discuss spin waves are:
\begin{align}
\partial_{t}f+\frac{\mathbf{p}}{m}\cdot\nabla_{\mathbf{r}}f-\nabla
_{\mathbf{r}}U_{0}\cdot\nabla_{\mathbf{p}}f-\nabla_{\mathbf{r}}\mathbf{U}%
\cdot\nabla_{\mathbf{p}}\mathbf{M} &  \simeq-\frac{f-f^{eq}}{\tau}\nonumber\\
\partial_{t}\mathbf{M}+\frac{\mathbf{p}}{m}\cdot\nabla_{\mathbf{r}}%
\mathbf{M}-\nabla_{\mathbf{r}}U_{0}\cdot\nabla_{\mathbf{p}}\mathbf{M}%
-\nabla_{\mathbf{r}}\mathbf{U}\cdot\nabla_{\mathbf{p}}f-\frac{2\mathbf{U}%
}{\hbar}\times\mathbf{M} &  \simeq-\frac{\mathbf{M}-\mathbf{M}^{eq}}{\tau
}\label{kinetic2}%
\end{align}
The main differences with the LL kinetic equation derived in Ref. \cite{LL1}
are that the full mean field is included (not just the spin mean field), and
that the equation is obtained for general spin-dependent interactions.
Nevertheless, our result is also less general, since the collisions are
assumed to occur only in the $s$-wave channel and that the collision integral
is treated at the relaxation time approximation level (if necessary, these two
restrictions could be lifted without any special difficulty).

\subsection{Mean field}

\label{mfa}

Another way to derive the modified drift term is to directly use a mean field
approximation within field theory. The following second quantized Hamiltonian
density is used to describe a trapped ultra-cold gas of bosons or fermions:
\begin{align}
\mathcal{H}  &  =\mathcal{H}_{1}+\mathcal{H}_{int}\nonumber\\
&  =\sum_{\alpha=1,2}\left(  \frac{\hbar^{2}}{2m}\nabla\psi_{\alpha}^{\dagger
}(\mathbf{r})\cdot\nabla\psi_{\alpha}(\mathbf{r})+V_{\alpha}^{ext}%
(\mathbf{r})\psi_{\alpha}^{\dagger}(\mathbf{r})\psi_{\alpha}(\mathbf{r}%
)\right) \nonumber\\
&  +\frac{1}{2}\sum_{\alpha,\beta}g_{\alpha\beta}\psi_{\alpha}^{\dagger
}(\mathbf{r})\psi_{\beta}^{\dagger}(\mathbf{r})\psi_{\beta}(\mathbf{r}%
)\psi_{\alpha}(\mathbf{r}) \label{ham}%
\end{align}
The annihilation and creation field operators of a particle in the internal
state $\alpha$ are $\psi_{\alpha}(\mathbf{r})$ and $\psi_{\alpha}^{\dagger
}(\mathbf{r})$. They have bosonic or fermionic equal-time commutation or
anticommutation relations. The interactions only depend on the three coupling
constants $g_{11}$, $g_{12}$ and $g_{22}$; the real potential $V$-matrix
elements have been replaced by the appropriate $T$-matrix elements at
low-energy, according to an usual procedure in the study of cold gases (Fermi
pseudo-potential method \cite{Huang}, or $V\rightarrow T$ renormalization
procedure in Ref. \cite{LPL9}).

We wish to obtain the time evolution of the single-particle density matrix
$\rho_{\alpha\beta}(\mathbf{r},\mathbf{r^{\prime}},t)=\langle\psi_{\alpha
}^{\dagger}(\mathbf{r^{\prime}},t)\psi_{\beta}(\mathbf{r},t)\rangle$, where
$\langle\ldots\rangle$ denotes the expectation value with respect to the
time-dependent $N$-body density operator of the system. To this end we write
down the Heisenberg equation of motion for $\psi_{\alpha}^{\dagger
}(\mathbf{r^{\prime}},t)\psi_{\beta}(\mathbf{r},t)$ using the Hamiltonian
(\ref{ham}). The non-interacting part of the Hamiltonian density
$\mathcal{H}_{1}$ gives the single-particle Liouville-von Neumann equation:
\begin{align}
i\hbar\partial_{t}\rho_{\alpha\beta}(\mathbf{r},\mathbf{r^{\prime}},t)  &
=-\frac{\hbar^{2}}{2m}\left(  \nabla_{\mathbf{r}}^{2}-\nabla_{\mathbf{r}%
^{\prime}}^{2}\right)  \rho_{\alpha\beta}(\mathbf{r},\mathbf{r^{\prime}%
},t)\nonumber\\
&  +\left(  V_{\alpha}^{ext}(\mathbf{r})-V_{\beta}^{ext}(\mathbf{r^{\prime}%
})\right)  \rho_{\alpha\beta}(\mathbf{r},\mathbf{r^{\prime}},t) \label{liouv}%
\end{align}
The interacting part $\mathcal{H}_{int}$ couples the single-particle density
matrix with the expectation value of the product of four field operators
(two-particle density matrix). If we are only interested in time scales large
compared to the duration of a collision, we can decompose the averages of such
products into averages of products of two operators \footnote{This result can
be obtained for instance from the Wick theorem.\ It is valid because the gas
is not Bose condensed; otherwise, the mean field could not be obtained in that
way, due to the absence of the exchange term.}. In the context of the equation
of motion for the single-particle density matrix, this approximation is often
called the random phase approximation (RPA) \cite{SJ}. The interacting part of
the equation of motion then becomes:
\begin{equation}
\sum_{\delta=1,2}V_{\alpha\delta}^{mf}(\mathbf{r},t)\rho_{\delta\beta
}(\mathbf{r},\mathbf{r^{\prime}},t)-\sum_{\delta=1,2}\rho_{\alpha\delta
}(\mathbf{r},\mathbf{r^{\prime}},t)V_{\delta\beta}^{mf}(\mathbf{r^{\prime}},t)
\label{meanfield}%
\end{equation}
where the mean field potential $2\times2$ matrix is given by:
\begin{equation}
V_{\alpha\beta}^{mf}(\mathbf{r})=\delta_{\alpha,\beta}\sum_{\gamma
=1,2}g_{\alpha\gamma}\langle\psi_{\gamma}^{\dagger}(\mathbf{r})\psi_{\gamma
}(\mathbf{r})\rangle+\epsilon g_{\alpha\beta}\langle\psi_{\alpha}^{\dagger
}(\mathbf{r})\psi_{\beta}(\mathbf{r})\rangle\label{veff}%
\end{equation}
which contains a direct term and an exchange mean field term, proportional to
$\epsilon$. Within this approximation, each particle evolves under the
influence of an effective single-particle potential $\hat{U}(\mathbf{r},t)$,
which is the sum of the external potential $\widehat{V}^{ext}(\mathbf{r})$ and
the time-dependent mean field $\hat{V}^{mf}(\mathbf{r},t)$.

In order to compare this result with that of the preceding section, we note
that $n_{\alpha}(\mathbf{r},t)=\langle\psi_{\alpha}^{\dagger}(\mathbf{r}%
,t)\psi_{\alpha}(\mathbf{r},t)\rangle$ is the average density of atoms in
internal state $\alpha=1,2$ and the off-diagonal elements $n_{12}%
(\mathbf{r},t)=n_{21}^{\ast}(\mathbf{r},t)=\langle\psi_{1}^{\dagger
}(\mathbf{r},t)\psi_{2}(\mathbf{r},t)\rangle$ are the coherences between the
internal states, so that the effective potential $\hat{U}$ reproduces exactly
the one obtained in the preceding section, equation (\ref{U}). We therefore
arrive at the following operatorial kinetic equation:
\begin{equation}
\frac{d\hat{\rho}_{1}}{dt}+\frac{1}{i\hbar}~\left[  \hat{\rho}_{1},\hat{H}%
_{1}+\hat{V}^{mf}\right]  _{-}=0 \label{QuantumVlasov}%
\end{equation}
which is the quantum equivalent of the Vlasov equation (see Ref.
\cite{Zubarev} for example); in this equation, $\hat{H}_{1}$ is the single
particle hamiltonian defined in (\ref{3}) and $\hat{V}^{mf}$ is defined by
(\ref{veff}). Obviously, if we Wigner transform this operatorial equation, we
will obtain a kinetic equation that is equivalent to (\ref{kinetic}), but
without the collision integral.\ The effect of lateral collisions can then be
treated phenomenologically by adding by hand a term corresponding to a
relaxation time approximation, as usual \cite{SJ}; then equations
(\ref{kinetic2}) are recovered and the mean field treatment of interactions
becomes exactly equivalent to these simplified equations.

\subsection{Discussion}

We have used three different methods to derive a kinetic equation for a gas of
atoms with two internal levels: the $S$ matrix (or LL) Ansatz, the Yvon-Snider
Ansatz (YS) and the mean field approximation; we now summarize a comparison
between the results.

To zero order in the gradients, LL and YS give exactly the same equation,
containing a Boltzmann-type collision integral and a spin mean field. As for
the mean field approximation, it provides only the spin mean field, obtained
in the limit of low-energy collisions because of the use of the
pseudopotential; this limit is in full agreement with the results of the two
other methods. We note in passing that this spin mean field occurs for a
non-condensed Bose gas, but not in a Bose-Einstein condensate (BEC) at very
low temperature, because the exchange term is absent when a single quantum
state is involved. In this case, there is however a different mechanism for
creating spin waves, the so-called \textquotedblleft quantum
torque\textquotedblright\ of purely kinetic origin \cite{Nikuni-Williams}; it
arises naturally because, in quantum mechanics, any gradient of the phase
corresponds to a kinetic energy for each component which, in turn, affects the
evolution of the relative phase and therefore of the transverse spin orientation.

To first order in the gradients and for forward scattering, LL, YS
and the mean field approximation all give the same scalar mean
field term (zero sound), again in the limit of low-energy
collisions for the third method. We have checked that the complete
expression of the LL and YS first order gradient terms coincide
exactly for forward scattering, but not for lateral scattering;
whether or not the two methods are completely equivalent to this
order is left as an open question; for more details see
\cite{theseJN}.

In a non-degenerate dilute gas, the spin mean field generally dominates over
the scalar mean field, because it appears to lower order in the gradient
expansion; it is therefore not surprising that zero sound type collective
modes should not propagate as easily as spin waves in dilute non-degenerate gases.

To what extent collisional methods and mean field approximations are
equivalent is not a trivial question.\ Generally speaking, mean field theory
is assumed to be valid for systems with either long range interactions
(Vlasov's equation for a plasma \cite{Zubarev}) or in strongly interacting
degenerate systems described in terms of quasi-particles (Landau's equation
for a Fermi liquid \cite{Kinetics,Silin,LPL9}); for a dilute gas, neither of
these conditions is met. Our conclusion is nevertheless that the equivalence
is perfect to first order in the scattering length $a$ but does not hold to
higher orders in $a$.\ We finally remark that, when describing a
non-degenerate dilute atomic gas in terms of quasi-particles (see the work of
Bashkin \cite{Bashkin2}, for example), a quasi-particle is just a particle
whose kinetic energy $p^{2}/2m$ is shifted by a local mean field term, so that
the kinetic drift term of the quasi-particle transport equation is left unchanged.

\section{Spin oscillations and internal conversion}

Our purpose now is to use the kinetic equation obtained in the preceding
section to discuss spin waves in ultra-cold trapped atomic gases. We first
briefly review a recent experiment realized at JILA \cite{Cornell} and the
theoretical work stimulated by it \cite{Oktel,Fuchs,Williams,Nikuni,Bradley}.
We then discuss the non-hydrodynamic character of the observed spin waves,
which allows a comparison between the different theoretical treatments.
Finally, we describe the decay of the spin waves and discuss the relevance of decoherence.

\subsection{The JILA experiment}

\label{JILA}In a recent experiment performed at JILA , Lewandowsky \emph{et
al} have studied bosonic $^{87}$Rb atoms with two hyperfine states of interest
(denoted by $1$ and $2$), confined in an axially symmetric magnetic trap
elongated along the $Ox$ direction.\ Their experiment is described in
\cite{Cornell}; the temperature $T$ of the gas in this experiment is about
twice the critical temperature for Bose-Einstein condensation, so that the gas
is not strongly degenerate and can be treated reasonably well as a Boltzmann
gas; however, since the de Broglie thermal wavelength is much larger than the
average scattering length, collisions occur in the full quantum regime.
Initially, the gas is at equilibrium with only state $1$ populated. A $\pi/2$
radio frequency pulse is then applied, which suddenly puts all the atoms into
the same coherent superposition of states $1$ and $2$. The subsequent
evolution of the system along the axial direction is then monitored by
measuring optically the local densities $n_{1}$ and $n_{2}$ of atoms in each
internal state. Experimentally, one observes that the system \textquotedblleft
segregates\textquotedblright: after about $100$~ms, atoms in state $1$ are
mostly found away from the center of the trap, while atoms in state $2$ move
towards the center, the total density $n=n_{1}+n_{2}$ remaining practically
unchanged. After about $200$~ms, the local densities of each species return to equilibrium.

In Ref. \cite{Fuchs}, this phenomenon was explained in terms of an internal
conversion resulting from the identical spin rotation effect (ISRE), which is
at the origin of the spin mean field; similar considerations were almost
simultaneously provided by two other groups \cite{Oktel,Williams}.
Qualitatively, this spin oscillation can be understood as follows:

(i) the field gradient creates an inhomogeneous spin precession, so that the
gas develops a gradient of transverse spin polarization

(ii) the thermal motion of the atoms then creates correlations between
transverse spin and velocity; therefore, a particle moving with a given
velocity at point $x$ gets a spin polarization which is not parallel to the
average local polarization

(iii) the ISRE then makes its spin polarization leave the transverse plane and
get a longitudinal component, in a direction which depends on the sign of the velocity

(iv) this is equivalent to a velocity dependent internal conversion which, at
some later time, results in an efficient spatial separation of atoms in the
two internal states.

To study this phenomenon quantitatively, we derive from (\ref{kinetic2}) an
effective one-dimensional kinetic equation in terms of the local density in
phase space $f(x,p,t)$ and spin density $\mathbf{M}(x,p,t)$, by assuming that
radial equilibrium is quickly established (the radial trap frequency
$\omega_{rad}/2\pi$ is much larger than the axial trap frequency $\omega
_{ax}/2\pi$). Integrating (\ref{kinetic2}) over radial coordinates $y,z$ and
momenta $p_{y},p_{z}$ provides the following equations:
\begin{align}
\partial_{t}f+\frac{p}{m}\,\partial_{x}f-\partial_{x}U_{0}\,\partial
_{p}f-\partial_{x}\mathbf{U}\cdot\partial_{p}\mathbf{M}  &  \simeq
-(f-f^{eq})/\tau\label{kinetic3f}\\
\partial_{t}\mathbf{M}+\frac{p}{m}\,\partial_{x}\mathbf{M}-\partial_{x}%
U_{0}\,\partial_{p}\mathbf{M}-\partial_{x}\mathbf{U}\,\partial_{p}f  &
-\frac{2\mathbf{U}}{\hbar}\times\mathbf{M}\nonumber\\
&  \simeq-(\mathbf{M}-\mathbf{M}^{eq})/\tau\label{kinetic3}%
\end{align}
(every quantity depending initially on three-dimensional coordinates
$\mathbf{r}$ and momenta $\mathbf{p}$ has been integrated over radial
coordinates and momenta, so that it now depends only on the coordinate $x$ and
the momentum $p\equiv p_{x}$). In the process of radial averaging, the
coupling constants $g_{\alpha\beta}$ are renormalized by a factor $1/2$, as
discussed by Levitov \cite{Oktel}. The three-dimensional effective trapping
potential (\ref{trappot}) therefore becomes:
\begin{align}
U_{0}(x,t)  &  =\frac{V_{2}^{ext}+V_{1}^{ext}}{2}+\left[  \frac{1+\epsilon}%
{2}(g_{22}+g_{11})+g_{12}\right]  \frac{n}{4}\nonumber\\
&  +\frac{1+\epsilon}{2}(g_{22}-g_{11})\frac{m_{\parallel}}{4}
\label{trappot1}%
\end{align}
and the three-dimensional effective magnetic field (\ref{magn}) is changed
into:
\begin{equation}
\mathbf{U}(x,t)=\frac{\hbar\Omega(x,t)}{2}\mathbf{e}_{\parallel}+\epsilon
\frac{g_{12}\mathbf{m}(x,t)}{4} \label{magn1}%
\end{equation}
with:
\begin{equation}
\hbar\Omega(x,t)=V_{2}^{ext}-V_{1}^{ext}+\frac{1+\epsilon}{2}\left[
(g_{22}-g_{11})\frac{n}{2}+(g_{22}+g_{11}-2g_{12})\frac{m_{\parallel}}%
{2}\right] \nonumber
\end{equation}
where $\mathbf{e}_{\parallel}$ is the unit vector in the longitudinal
direction in spin space.

A few simplifying assumptions are appropriate for the experimental conditions
of Ref. \cite{Cornell}. The confining energy $(V_{2}^{ext}+V_{1}^{ext})/2$ is
of order $k_{B}T\simeq13$~kHz$\times h$ which is much larger than the mean
field interaction energy $gn(0)\simeq140$~Hz$\times h$. This allows us to keep
only the confining energy of the harmonic trap in the effective trapping
potential $U_{0}$:
\begin{equation}
U_{0}(x)\simeq\frac{V_{1}^{ext}(x)+V_{2}^{ext}(x)}{2}=\frac{1}{2}m\omega
_{ax}^{2}x^{2}%
\end{equation}
The differential trapping energy $V_{2}^{ext}-V_{1}^{ext}\sim10$~Hz$\times h$
is even weaker than the mean field interaction energy. This ensures that the
force terms $\partial_{x}\mathbf{U}\cdot\partial_{p}(f\text{ or }\mathbf{M})$
appearing in equations (\ref{kinetic3f},\ref{kinetic3}) are negligible. On the
contrary, when the local effective magnetic field $\mathbf{U}$ does not appear
under a spatial gradient, as in the term $2\mathbf{U}\times\mathbf{M}/\hbar$,
it can not be neglected. This effective magnetic field $\mathbf{U}$ is made of
two terms (see equation (\ref{magn1})): one is an effective \emph{external}
magnetic field $\hbar\boldsymbol{\Omega}/2$; the other is an \emph{exchange}
magnetic field or spin mean field $g_{12}\mathbf{m}/4$ which results from the
ISRE. The effective external magnetic field is the sum of the contributions of
a differential Zeeman and of a differential mean field:
\begin{equation}
\hbar\boldsymbol{\Omega}(x)=\hbar~\Omega(x)\mathbf{e}_{\parallel}\simeq\lbrack
V_{2}^{ext}(x)-V_{1}^{ext}(x)+(g_{22}-g_{11})n(x)/2]\mathbf{e}_{\parallel}
\label{eemf}%
\end{equation}
where, following ref. \cite{Cornell}, we have assumed that $2g_{12}\simeq
g_{11}+g_{22}$ for simplicity. The average value over the sample of the
effective external magnetic field can be removed by going to a uniformly
rotating frame (Larmor frame). For the numerical simulations, we need to know
the $x$ dependence of this effective magnetic field. As the profile $n(x)$
will be shown to be a Gaussian (as for a non-interacting non-degenerate gas)
which does not vary in time, we model the experimentally measured effective
external magnetic field by:
\begin{equation}
\boldsymbol{\Omega}(x)=-\delta\Omega\exp{-(m\omega_{ax}^{2}x^{2}/2k_{B}%
T)}\mathbf{e}_{\parallel} \label{a}%
\end{equation}
where the parameter $\delta\Omega$ is the variation of $\Omega(x)$ between the
center and the edge of the cloud.

Taking into account all approximations mentioned above, we finally arrive at
equations:
\begin{align}
\partial_{t}f+\frac{p}{m}\,\partial_{x}f-m\omega_{ax}^{2}x\,\partial_{p}f  &
\simeq-(f-f^{eq})/\tau\label{kinetic4f}\\
\partial_{t}\mathbf{M}+\frac{p}{m}\,\partial_{x}\mathbf{M}-m\omega_{ax}%
^{2}x\,\partial_{p}\mathbf{M}  &  -(\boldsymbol{\Omega}+\frac{g_{12}%
\mathbf{m}}{2\hbar})\times\mathbf{M}\nonumber\\
&  \simeq-(\mathbf{M}-\mathbf{M}^{eq})/\tau\label{kinetic4}%
\end{align}
The initial equilibrium Maxwell-Boltzmann distribution $f(x,p)$ solves the
kinetic equation (\ref{kinetic4f}), so that the dynamics after the $\pi/2$
pulse can be expressed in terms of $\mathbf{M}$ only. In addition, the density
$n(x)$ is obtained by integrating the equilibrium Maxwell-Boltzmann
distribution over momentum $p$, which shows that it is time independent; from
now on, we will only consider equation (\ref{kinetic4}). The spin distribution
immediately after the $\pi/2$ pulse is assumed to be:
\begin{equation}
\mathbf{M}(x,p,t=0)=f(x,p)\mathbf{e}_{\perp,1}%
\end{equation}

\begin{figure}[ptb]
\begin{center}
\includegraphics[height=5cm]{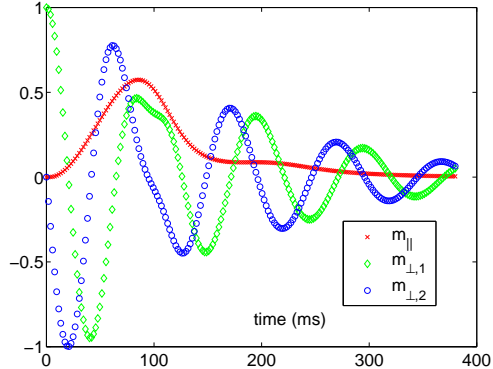}
\end{center}
\caption{Time evolution of the spin polarization $\mathbf{m}$ at the center of
the trap ($x=0$) when $\delta\Omega/2\pi=12$~Hz.}%
\label{mzt}%
\end{figure}

Equation (\ref{kinetic4}) can be solved numerically with values of the
parameters taken from Ref. \cite{Cornell}: the trap frequencies are
$\omega_{ax}/2\pi=7$~Hz and $\omega_{rad}/2\pi=230$~Hz; the average time
between collisions is $\tau\sim10$~ms; the temperature is $T\simeq0.6$~$\mu$K;
the density at the center of the trap is $n(0)=1.8\times10^{13}$~cm$^{-3}$;
$\delta\Omega/2\pi$ is typically $\sim12$~Hz, and the scattering lengths are
$a_{11}=100.9a_{0}$, $a_{22}=95.6a_{0}$ and $a_{12}=98.2a_{0}$ where $a_{0}$
is the Bohr radius. The solution at the center of trap is plotted in Figure
\ref{mzt}; it shows very good agreement with the experimental observations,
without any adjustable parameter. 

\subsection{Large, non-hydrodynamic, spin waves}

Ordinary spin waves in non-degenerate gases usually occur either in the
hydrodynamic regime or/and in a regime of small amplitudes
\cite{Hpol,Hepol,Gully}. One can then assume a small departure of the spin
distribution (from either local or global equilibrium); this is similar to the
first order Chapman-Enskog gradient expansion, which leads to the
Navier-Stokes equations (see for instance \cite{Huang}). The result is the
well known Leggett equations \cite{Leggett}, transposed from degenerate Fermi
liquids to dilute non-degenerate gases \cite{LL2}:%
\begin{equation}%
\begin{array}
[c]{cc}%
\partial_{t}\mathbf{m}+\partial_{x}\mathbf{j}=\boldsymbol{\Omega}\times
\mathbf{m} & \\
\partial_{t}\mathbf{j}-(\boldsymbol{\Omega}+\frac{g_{12}\mathbf{m}}{2\hbar
})\times\mathbf{j}+\frac{k_{B}T}{m}\partial_{x}\mathbf{m}+\omega_{ax}%
^{2}x\mathbf{m} & \simeq-\frac{\mathbf{j}}{\tau}%
\end{array}
\label{Leggett}%
\end{equation}
where the spin current along $Ox$ is defined by:
\begin{equation}
\mathbf{j}(x,t)\equiv\int dp~\frac{p}{m}\mathbf{M}(x,p,t)
\end{equation}
The momentum dependence has disappeared from these equations, which contain
only position variables. These equations can describe spin waves, not only in
the hydrodynamic regime, but also in the collisionless regime as, for example,
in liquid $^{3}$He (Silin spin waves \cite{Masuhara}) or in H$\downarrow$ gas
\cite{Bigelow}. For a discussion of the hydrodynamic-like description of spin
waves in the collisionless regime of non-degenerate dilute gases, see Refs.
\cite{Bashkin2,Meyerovich2}.

\begin{figure}[ptb]
\begin{center}
\includegraphics[height=4cm]{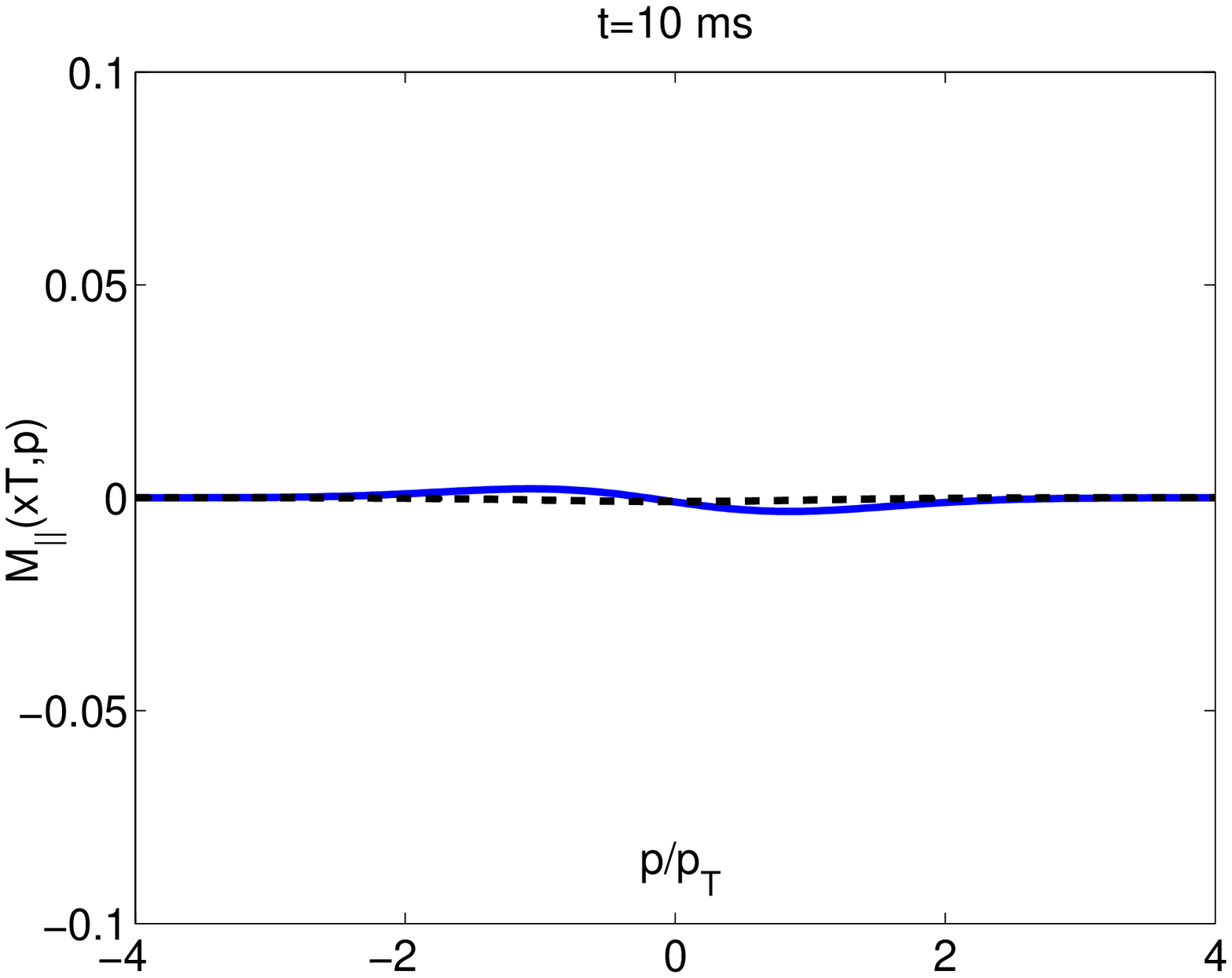} \includegraphics
[height=4cm]{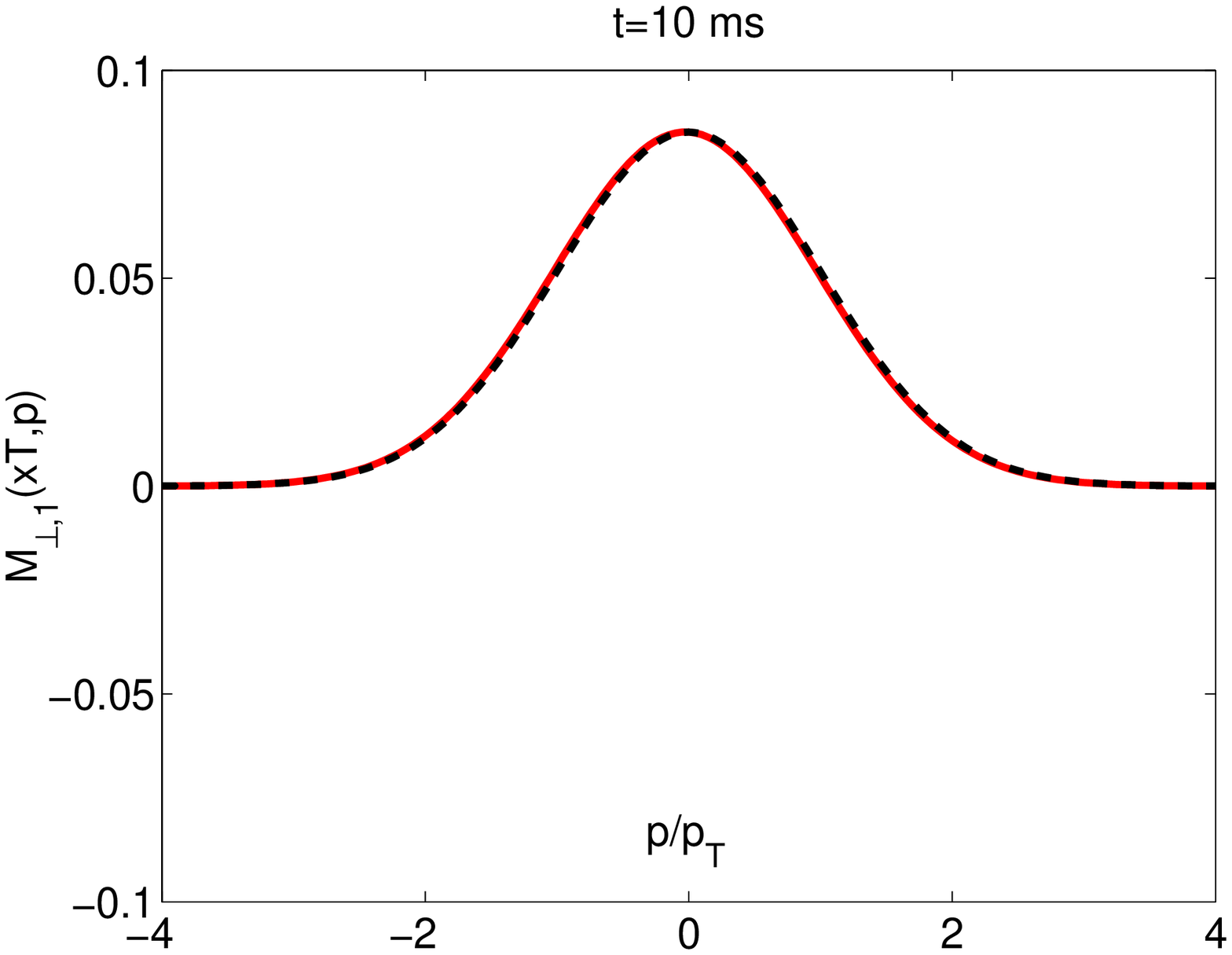}
\par
\includegraphics[height=4cm]{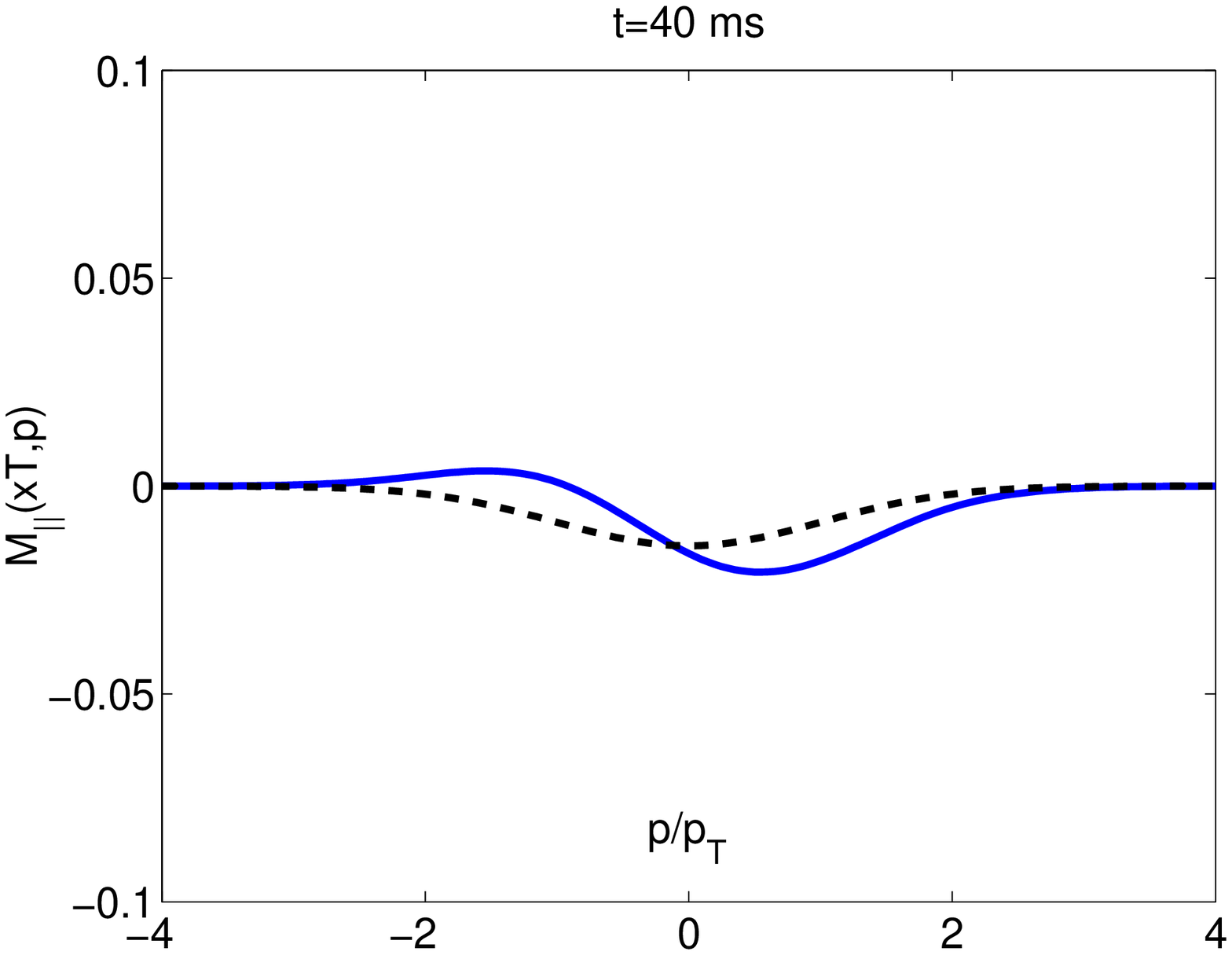} \includegraphics
[height=4cm]{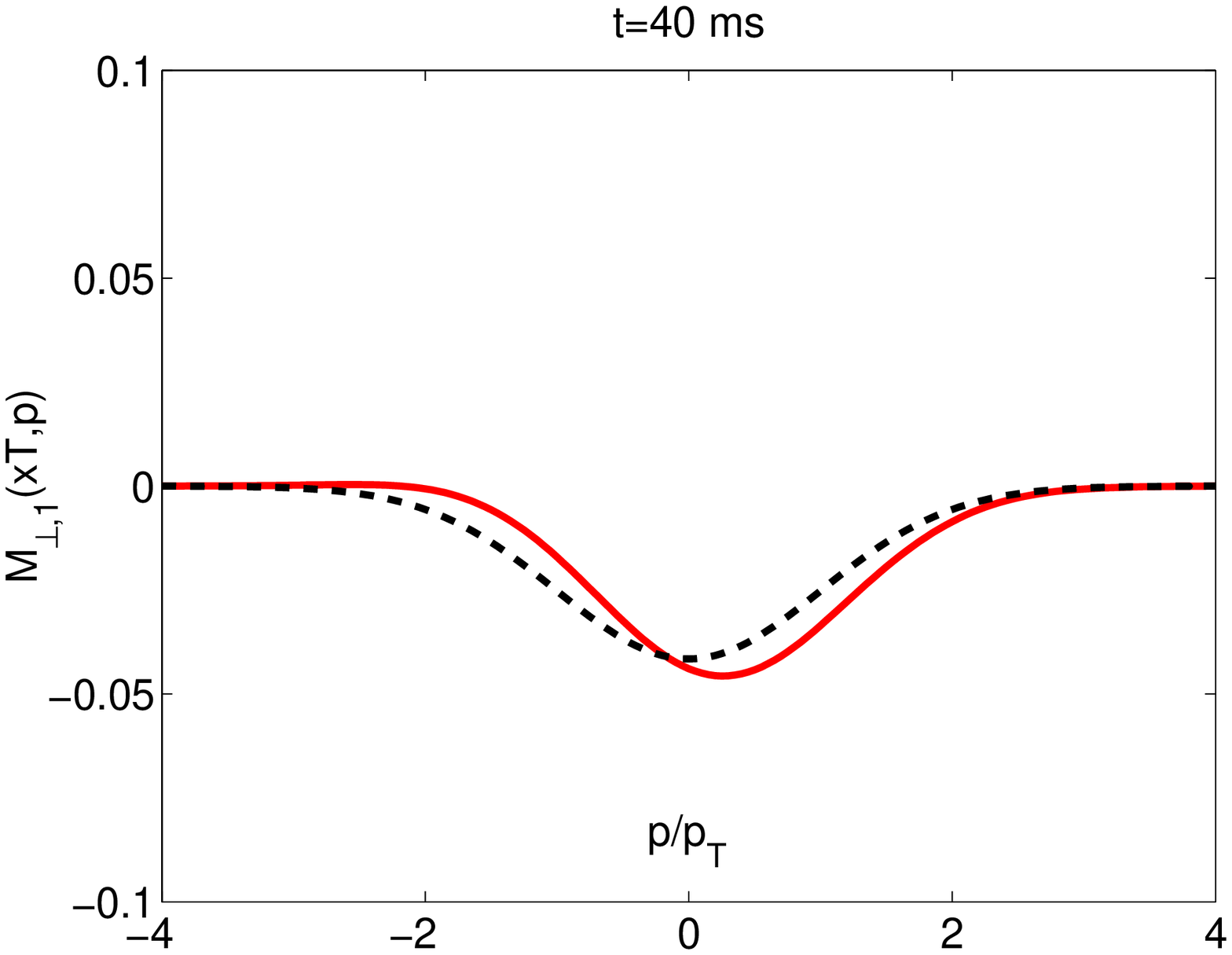}
\par
\includegraphics[height=4cm]{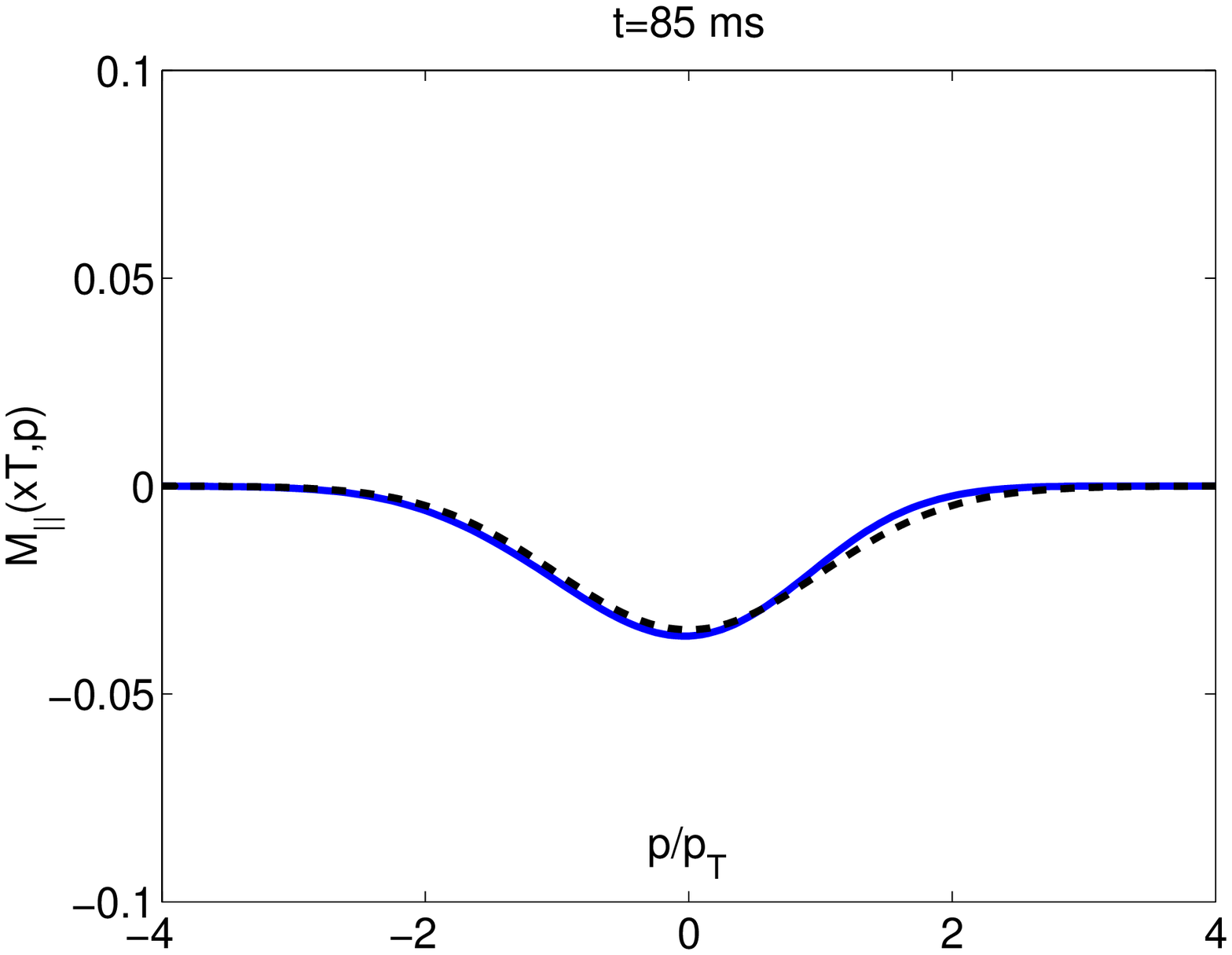} \includegraphics
[height=4cm]{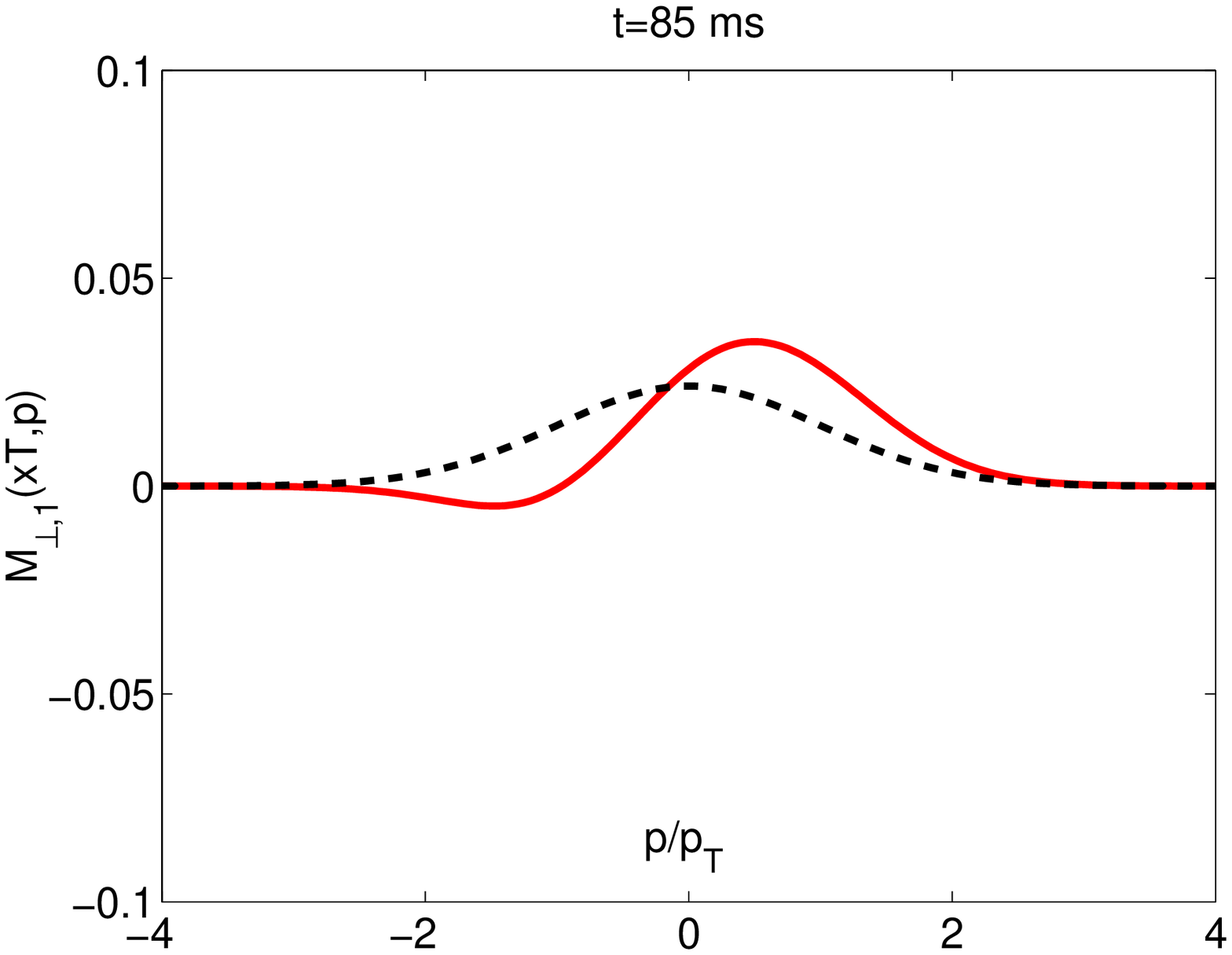}
\par
\includegraphics[height=4cm]{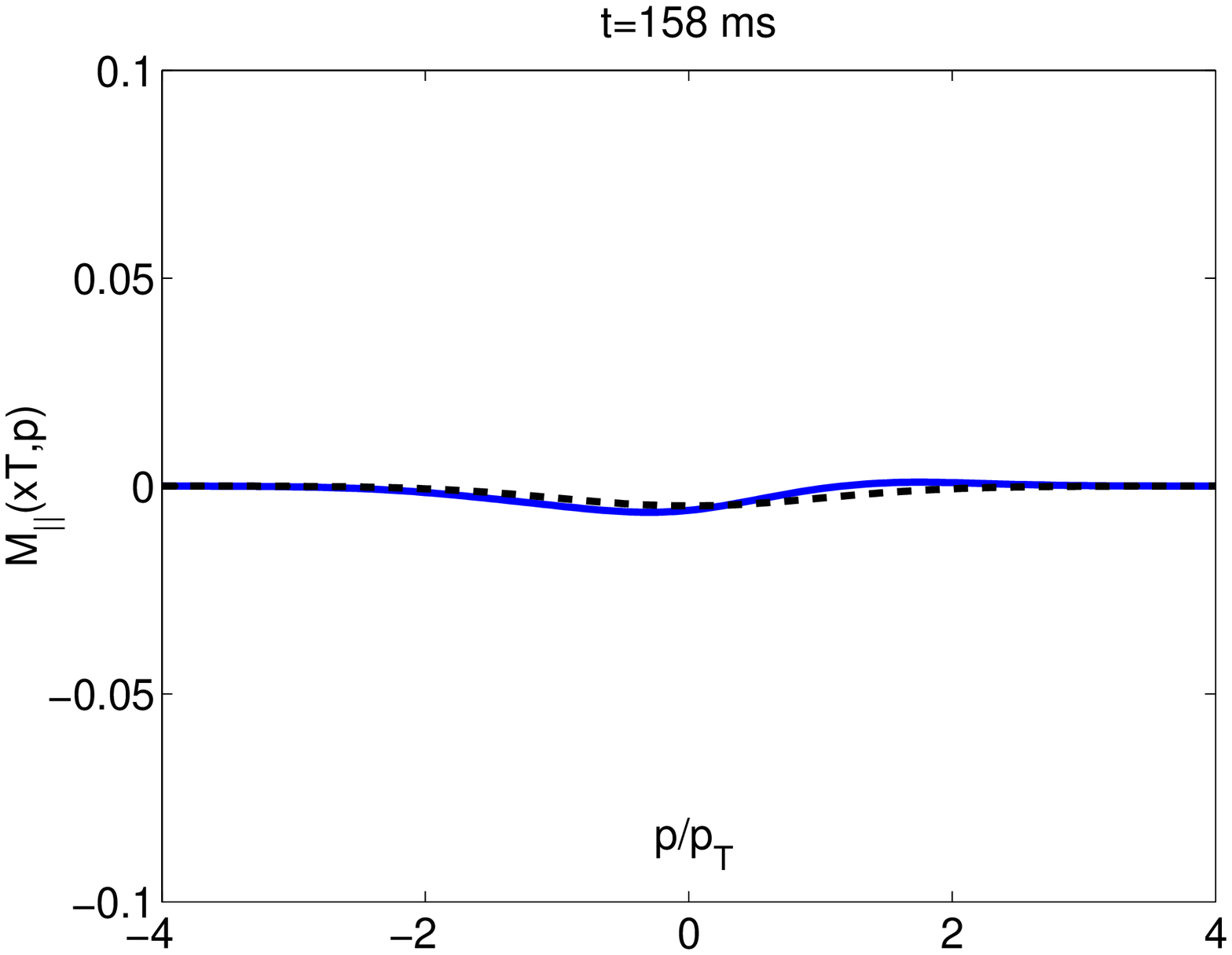} \includegraphics
[height=4cm]{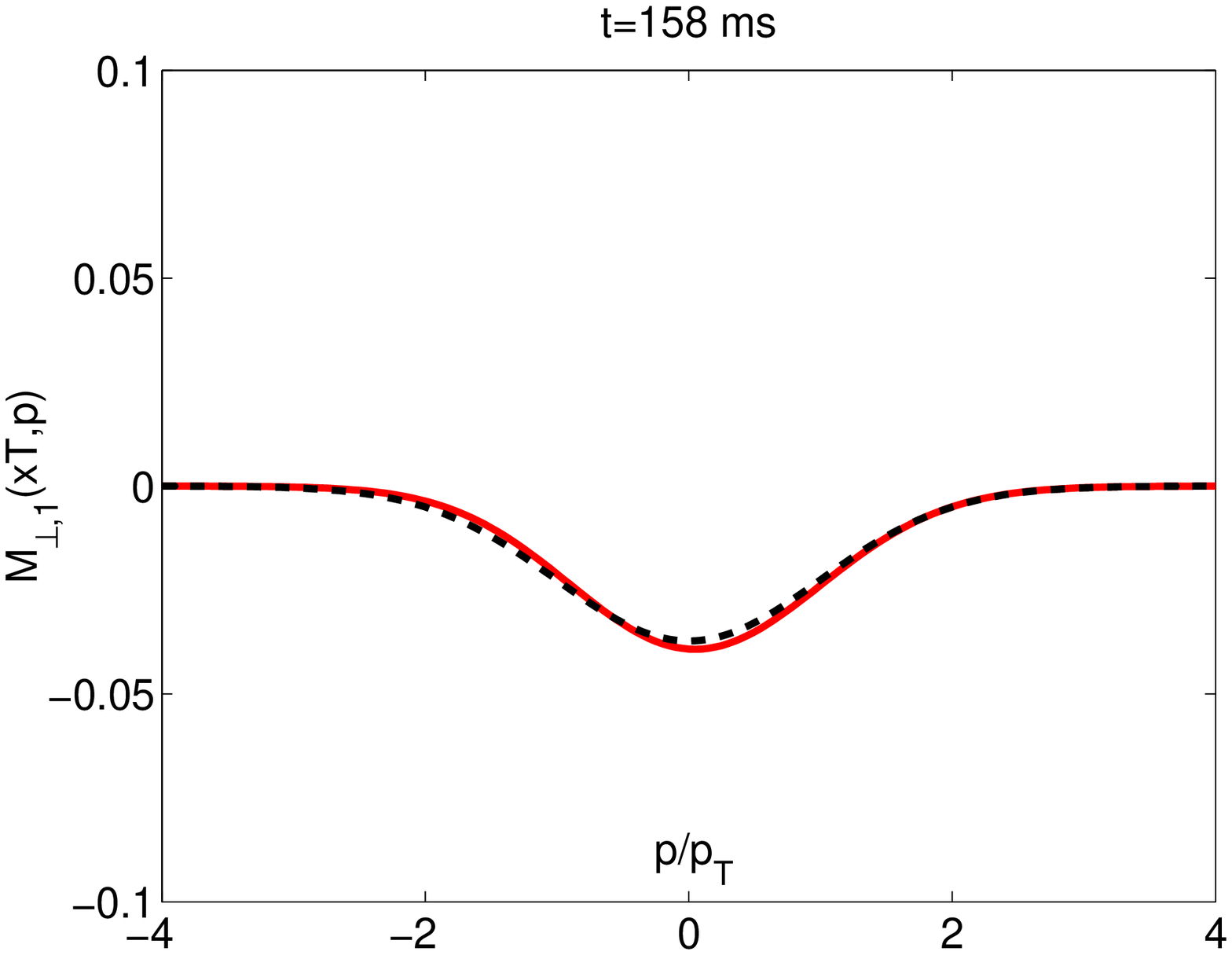}
\end{center}
\caption{Longitudinal spin distribution $M_{||}(x=x_{T},p,t)$ and transverse
spin distribution $M_{\bot,1}(x=x_{T},p,t)$ at position $x=x_{T}$ for
$t=10;40;85;160$~ms, where $x_{T}=\sqrt{k_{B}T/m\omega_{ax}^{2}}$ and
$p_{T}=\sqrt{mk_{B}T}$. The local equilibrium spin distribution is plotted
with a dashed line.}%
\label{pdistr}%
\end{figure}

The spin waves observed at JILA were of large amplitude and
occurred in the intermediate regime between hydrodynamic and
collisionless, where $\omega_{ax}\tau\sim1$, so that there is no a
priori reason to believe in the validity of the Leggett equations;
we therefore need to resort to a numerical solution of the kinetic
equation in terms of both position and momentum variables. Figure
\ref{pdistr} shows the results of this calculation, and the spin
density distribution (longitudinal and transverse) in $p$-space
for different times. For comparison, the local equilibrium spin
distribution is also plotted (dashed line); we see that the spin
distribution in $p$-space can indeed get very distorted, which
illustrates a strongly non-hydrodynamic situation.
For very short times, $t\lesssim40$~ms, the transverse and
longitudinal spin distributions are still close to local equilibrium
($M_{||}^{eq}=0$), so that the Leggett equations are valid. For $t\gtrsim
40$~ms, this is no longer true, so that one would lose a significant part of
the physics by not retaining the full momentum dependence of the distribution.
In other words, studying the two first moments of the distribution
($\mathbf{m}$ and $\mathbf{j}$) is not equivalent to studying the full
distribution $\mathbf{M}(p)$. What actually takes place is a phenomenon
analogous to a density \textquotedblleft shock wave\textquotedblright\ rather
than a \textquotedblleft sound wave\textquotedblright. Eventually, for
$t\gtrsim160$~ms (end of the \textquotedblleft segregation\textquotedblright),
since equilibrium is almost reached, the Leggett equations become valid again.

The initial part of the internal conversion (or state separation) can be
treated analytically in several approximations, allowing for comparison
between different approaches such as those of
Refs.\ \cite{Fuchs,Oktel,Williams}. In Ref. \cite{Fuchs}, we solved the
kinetic equation analytically, for times $t\ll\tau$, by expanding in Taylor
series in time and obtained:
\begin{equation}
\frac{m_{||}(x,t)}{n(x)}=\frac{g_{12}n(x)}{2\hbar}[\frac{k_{B}T}{m}%
\Omega^{\prime\prime}(x)-2\omega_{ax}^{2}x\Omega^{\prime}(x)]\frac{t^{4}}{4!}
\label{t4begin}%
\end{equation}
where $\Omega^{\prime}$ and $\Omega^{\prime\prime}$ are the first and second
spatial derivative of $\Omega(x)$. \begin{figure}[ptb]
\begin{center}
\includegraphics[height=4.15cm]{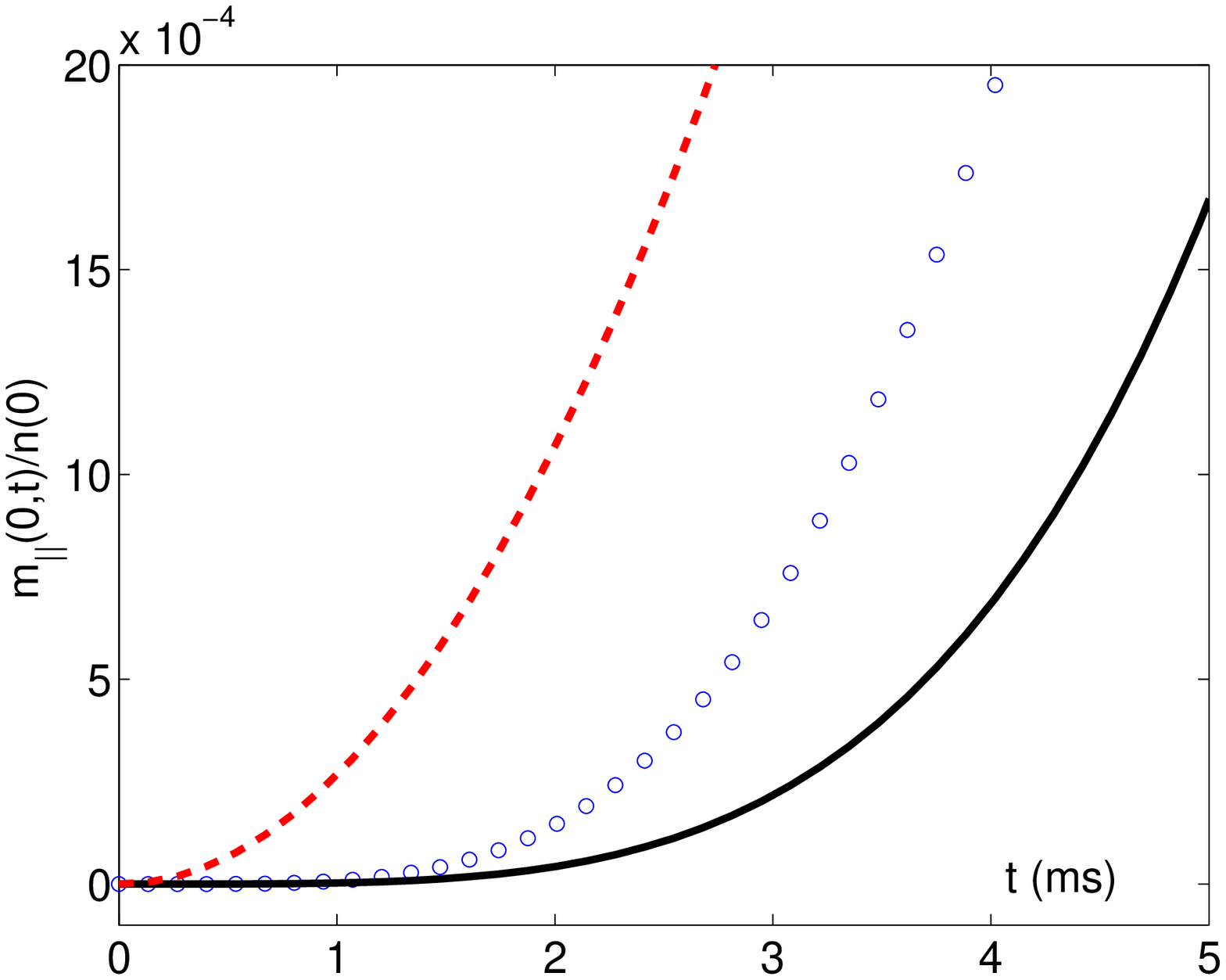} \includegraphics[height=4cm]{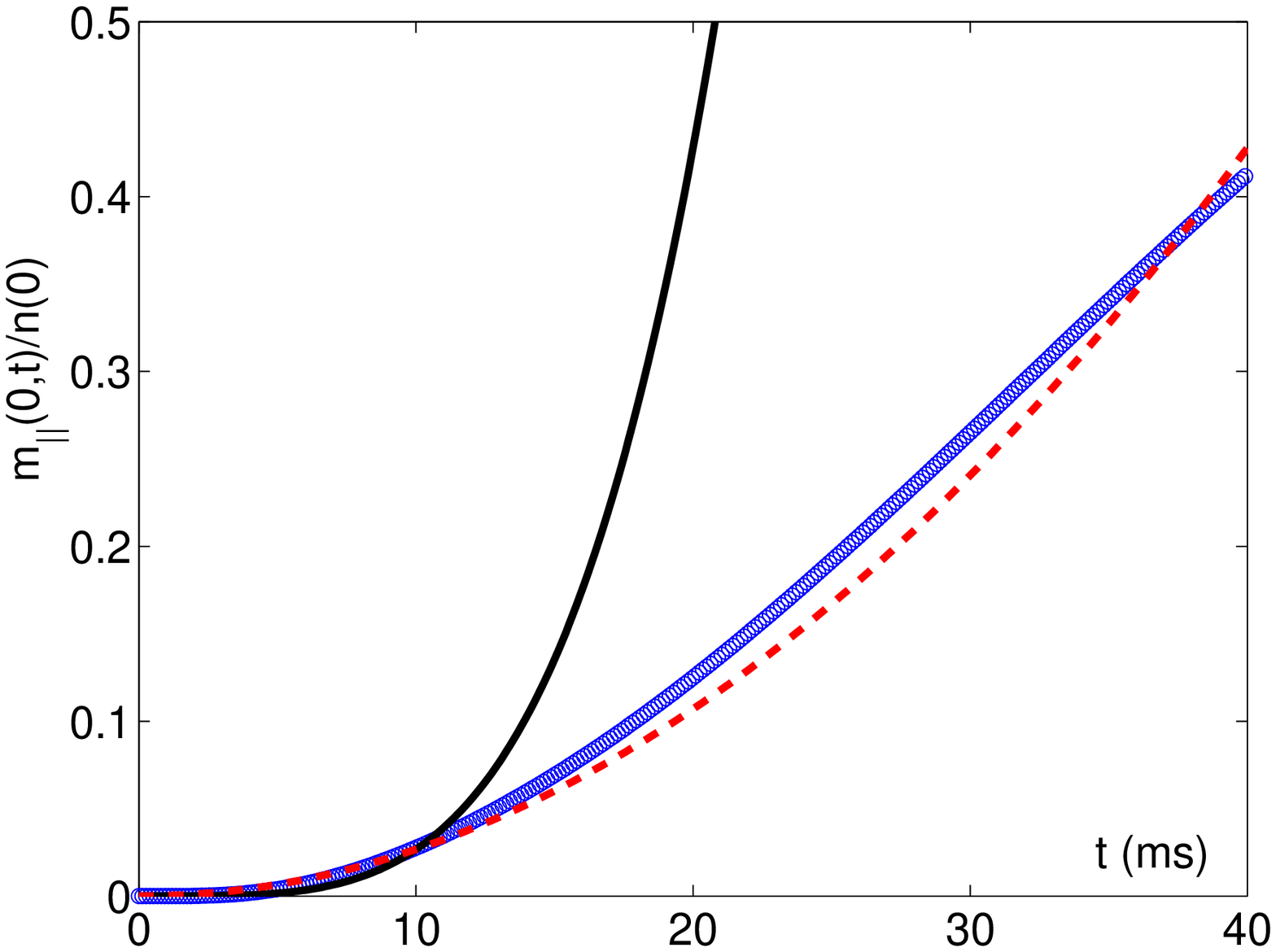}
\end{center}
\caption{Beginning of the internal conversion at the center of the trap. The
numerical simulation is plotted with circles, the $t^{4}$ result with a full
line and the $t^{2}$ result with a dashed line.}%
\label{t4}%
\end{figure}This formula predicts an initial quartic in $t$ behavior, which
correctly reproduces the numerical results; each power of $t$ corresponds to
one of the four physical processes (i) to (iv) described in \S \ \ref{JILA}.

Another approach to calculate the short time behavior is that used by Williams
\emph{et al} \cite{Williams}. They use the Leggett equations which are valid
for $t\ll\tau$ because the spin distribution is at equilibrium at $t=0$.
Solving the Leggett equations in the small time limit, where $-\mathbf{j}%
/\tau$ is negligible, they recover (\ref{t4begin}).

\begin{figure}[ptb]
\begin{center}
\includegraphics[height=5cm]{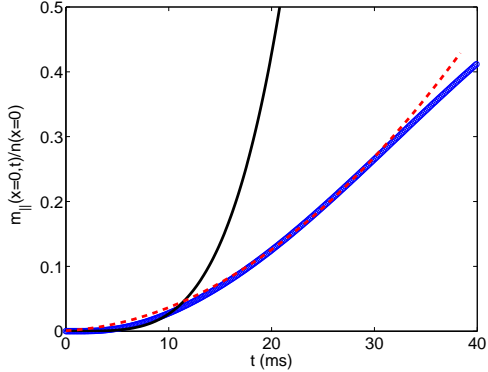}
\end{center}
\caption{The numerical simulation is plotted with circles, the $t^{4}$ result
with a full line and the $(t-t_{0})^{2}$ result with a dashed line ($t_{0}%
\sim2$~ms here). Note the good agreement between the $(t-t_{0})^{2}$ law and
the kinetic equation in the range $\tau\simeq10$~ms$<t\lesssim40$~ms.}%
\label{t2modt}%
\end{figure}

A third approach is to consider, following Oktel and Levitov
\cite{Oktel}), the Leggett equations in the hydrodynamic regime.
The approximation consists in assuming that the spin current
remains close to its stationary value, which
allows to neglect the term $\partial_{t}\mathbf{j}$ in equation (\ref{Leggett}%
). Using the resulting equations, we obtain a different behavior for small
times:
\begin{equation}
\frac{m_{||}(x,t)}{n(x)}\sim\frac{g_{12}n(x)}{2\hbar}[\frac{k_{B}T}{m}%
\Omega^{\prime\prime}(x)-2\omega_{ax}^{2}x\Omega^{\prime}(x)]\frac{(\tau
t)^{2}}{4!} \label{t2begin}%
\end{equation}
(the effect of the hydrodynamic approximation is to replace
$t^{2}$ by $t\tau $). The $t^{4}$ and $t^{2}$ results are plotted
in Fig. \ref{t4}; we see that times smaller than a time between
collisions $\tau$ can not be dealt with in this approach.
Nevertheless, after each atom has made one collision on average,
the spin current is close to its stationary value and the
hydrodynamic approximation is reasonably valid. Therefore, when
$t>\tau \simeq10$~ms, we get:
\begin{equation}
\frac{m_{||}(x,t)}{n(x)}\sim\frac{g_{12}n(x)}{2\hbar}[\frac{k_{B}T}{m}%
\Omega^{\prime\prime}(x)-2\omega_{ax}^{2}x\Omega^{\prime}(x)]\frac{\tau
^{2}(t-t_{0})^{2}}{4!} \label{t2modt-eq}%
\end{equation}
(the parameter $t_{0}$ includes the accumulated effect of the retardation of
the spin current $\mathbf{j}$ for small time). The $(t-t_{0})^{2}$ behavior is
plotted in Fig. \ref{t2modt} (the reason why, after $40$~ms, the numerical
solution of the kinetic equation is not well approximated by the
$(t-t_{0})^{2}$ law is that the spin distribution gets very distorted, so that
the Leggett equations are no longer valid). 

\subsection{Decay of the phenomenon}

Up to this point, we have studied mostly the initial part of the phenomenon,
when the amplitude of the spin wave grows and the two species tend to separate
from each other; we now discuss its final part, when the spin oscillations
decay and, eventually, the system returns to equilibrium. A first remark is
that it is relatively easy to obtain an order of magnitude of the time at
which the transition between the two regimes occurs, i.e. the time at which
the maximum of the separation takes places. For short times, we have seen that
this separation is an indirect result of a differential precession of the
transverse component of the spin at different points along the $Ox$ axis; the
thermal motion of the atoms then creates correlations between the transverse
orientation of the spins and the sign of the $x$ component of the
velocities.\ But, for long times, the differential rotation of the spins will
be so large that each sign of this velocity will become associated with a
widely open fan of transverse orientation, with almost zero average; clearly,
the selective internal conversion effect will then also average to almost
zero.\ Let us consider the simple situation of a homogeneous gas in a box of
size $L$ with a linear gradient of external effective magnetic field
$\Omega^{\prime}$. We focus, for instance, on the situation at the center of
the box. At time $t$, atoms initially at $x=v\times t$ will cross this point
with a transverse orientation rotated by an angle of the order of:
\begin{equation}
\Omega^{\prime}\times x\times t=\Omega^{\prime}~v~t^{2}%
\end{equation}
Therefore, at times greater than:
\begin{equation}
t_{m}\sim\frac{1}{\sqrt{\Omega^{\prime}~v}}%
\end{equation}
the transverse directions of the spins average out so that the apparent
segregation effect no longer takes place. In the case of the trapped gas,
$\Omega^{\prime}\sim\delta\Omega/L$ and $v\sim\omega_{ax}L$ where $L$ is the
size of the cloud, and we obtain:
\begin{equation}
t_{m}\sim\frac{2\pi}{\sqrt{\delta\Omega~\omega_{ax}}}%
\end{equation}
Typical values give $t_{m}\sim100$~ms in accordance with the results of the
simulation. After this maximum, the system tends to return to equilibrium
during a time period which we call the decay of the phenomenon.

Fig.\ref{mzt} shows that the decay of the longitudinal and
transverse spin polarizations are significantly different. The
longitudinal spin polarization $m_{\parallel}(x,t)$ returns to
equilibrium in two steps: it first quickly decreases from its
maximum value to almost zero in a time scale of the order of
$50$~ms; it then goes to zero much more slowly, on a time scale of
the order of $200$~ms. In order to understand this, we note in
Fig.\ref{mzt} that, at the maximum of separation, the wave packets
(or the clouds) associated with each internal state overlap only
little; the first step is then easily interpreted as the free
motion of the clouds under the restoring force of the trap. It
takes a quarter of a period $T_{ax}/4=\pi/2\omega _{ax}=36$~ms to
go from the maximum of oscillation to the center of the trap.
Actually, as soon as the two clouds start to overlap, they
interact through a repulsive mean field ( $g>0$), so that it takes
a little more time than a quarter of a period to reach the
situation where the two clouds overlap significantly
\footnote{That the two clouds interact mainly trough the mean
field and not trough (lateral) collisions is confirmed by the
numerical solution of the kinetic equation. Indeed, we checked
that removing the collision integral does not change the decay
time scale. The role of the collision integral is merely to damp
revivals of the longitudinal spin waves, which are observed when
the collision integral is discarded.}. At this point, the
subsequent evolution of the system can be understood as the mutual
diffusion of two gases \footnote{Actually, an instability of the
transverse spin polarization is possible when there is a strong
longitudinal spin polarization gradient (which is the case when
the two clouds overlap again)
and a spin mean field. This is known as Castaing's instability \cite{Castaing}%
. In Ref. \cite{Fuchs2}, we checked that it does not play a role in the
experiment done at JILA \cite{Cornell}.}; this happens $\sim150$~ms after the
initial pulse and can again be described by the Leggett equations (we checked
on the $p$ distribution that the system is close to local equilibrium). The
relaxation time scale is then given by the spin diffusion time $t_{diff}\sim
L^{2}/D$ (where $L\sim2\sqrt{k_{B}T/m\omega_{ax}^{2}}$ is the size of the
cloud and $D=k_{B}T\tau/m$ is the spin diffusion constant). Using the same
parameters as in the simulation, we obtain $t_{diff}\sim200$~ms, in good
agreement with the numerical result.

We now discuss the transverse spin polarization $\mathbf{m}_{\perp}(x,t)$. As
seen on Fig.\ref{mzt}, it oscillates many times before going to zero. After
$\sim150$~ms, when the Leggett equations are valid again, we know \cite{LL2}
that spin waves occur in the hydrodynamic regime. The spectrum derived from
these equations predicts damped transverse spin waves \cite{LL2}. The damping
time is of the order of the spin diffusion time $t_{diff}\sim200$~ms, which
gives an order of magnitude for the time it takes the transverse spin
polarization to go from its value at the maximum of separation to zero. This
is in agreement with the relaxation time obtained by numerically solving the
kinetic equation (Fig.\ref{mzt}). We finally note that the frequency of the
transverse spin waves is essentially given by the external effective magnetic
field precession frequency $\delta\Omega/2\pi=12$~Hz.

\subsection{``Ghost" wave packet and decoherence}

We now come back to the time at which the separation is maximum. In order to
simplify the discussion, we consider two clouds of atoms, one corresponding to
state $1$ and located around $x=d/2$, the other corresponding to state $2$
located around $x=-d/2$, where $d$ is the distance between the center of the
two clouds; we note $\delta x$ the width of each of these wave packets,
assuming that $\delta x<d$. In this case, elementary quantum mechanics
predicts that the local transverse spin polarization vanishes everywhere (this
is because the local spin density corresponds to an operator which is local in
ordinary space). Nevertheless, the superposition of the two wave packets
remains coherent so that, rigorously speaking, the system is not equivalent to
a classical mixture of two gases, in opposition to what we have assumed in the
previous section.

One may wonder how this coherence translates in terms of the Wigner
distribution. The answer is well known: the presence of the coherence is
contained in a so-called \textquotedblleft ghost" wave packet, which exists
around the middle point between the two real wave packets (or clouds). For
spinless particles, it \textquotedblleft carries interferences with it"
\cite{S-B}, and reconstructs them as soon as the real packets overlap again.
For particles with spin, when the two separated wave packets have opposite
spin orientations, the ghost wave packet centered at $x=0$ appears only in the
non-diagonal spin matrix elements (transverse spin components of the Wigner
distribution). In both cases, the wave packet rapidly oscillates as a function
of momentum $p$, taking positive and negative values, with an oscillation
period $2\pi\hbar/d$; the further apart the two clouds, the faster the
oscillation. On the other hand, for a statistical mixture of the internal
states, the ghost wave packet does not exist. Such oscillating wave packets
are well-known in the context of macroscopic superpositions of states
\cite{S-B}.

The next question is to what extent the ghost wave packet is preserved by the
time evolution of the kinetic equation. Since the Wigner formalism in itself
implies no approximation (it is strictly equivalent to the use of operators),
the question arises only because our kinetic equation does not provide the
exact quantum evolution: it was actually obtained from the operatorial
(quantum) equation for the single-particle density operator by truncating the
gradient expansion of the Wigner transform.\ In other words, we made a
semi-classical approximation. But, for the ghost wave packet, the $p$
oscillation period is $2\pi\hbar/d$ while the width of the wave packet is
$\delta x$, so that the parameter of the gradient expansion is $d/\delta x$,
which may be larger than $1$: therefore, our kinetic equation does not
necessarily remain correct for this case; the semi-classical gradient
expansion may smooth the oscillations of the ghost wave packet, and
artificially introduce decoherence. Another related source of artificial
decoherence\footnote{Concerning the second source of artificial decoherence,
we think that it is not a problem in our simulations.\ We checked that, when
removing the collision integral so that the kinetic equation becomes
reversible, the transverse spin polarization disappears when the two wave
packets do not overlap but reappears when they mix again.} might come from the
discretization of phase-space (necessary for the numerical solution of the
kinetic equation), which may introduce a lattice spacing in momentum space
larger than $\delta p$.

Up to this point, no real physical decoherence mechanism was
included in our discussion: we have assumed that the system is
perfectly isolated. But this is not the case in practice.\ In
order to obtain an estimate of the physical decoherence time, we
proceed by analogy with a well-understood situation in quantum
optics: a macroscopic superposition of two coherent states of the
electro-magnetic field in a cavity. In that case, it is known that
the coherence is lost as soon as one photon on average escapes the
cavity and goes in the environment \cite{Raimond,Haroche}. We
therefore assume here that for our trapped gas, a source of
decoherence comes from the fact that atoms can leave the trap,
because of 3-body collisions for example. The time it takes one
atom to leave the trap is $1/\alpha_{3b}n(0)^{2}N$ where
$\alpha_{3b}\sim 4\times10^{-29}$~cm$^{6}$/s is the 3-body
recombination rate constant \cite{Cornell3}, $N\sim10^{6}$ is the
total number of atoms and $n(0)\simeq 2\times10^{13}$~cm$^{-3}$ is
the density at the center of the trap. This gives $\sim0.1$~ms as
a rough upper bound to the physical decoherence time, which is
still much shorter than the characteristic time scale of the spin
state segregation. Our conclusion is therefore that, when the wave
packets recombine, coherence is not likely to play a role in these
experiments; this justifies our discussion in the preceding
section.

\section{Conclusion}

We have derived a kinetic equation for a Boltzmann gas with two internal
levels using different methods. A first conclusion which emerges from this
work is that, for dilute gases at low temperature, several significantly
different theoretical approaches lead essentially to the same result.\ Here we
have discussed the Yvon-Snider method, with its peculiar way to close the
BBGKY hierarchy, the $S$ matrix method which is very close to the spirit of
the original Boltzmann equation, as well as the popular mean field method.\ In
the latter case, the equivalence actually holds only in the low temperature
limit, when forward scattering dominates over lateral scattering by a factor
of the order of $\lambda_{T}/a$.\ Whether or not it is possible to find
experimental situations where the results of these theories are significantly
different, and can be tested experimentally, is still an open question (for
instance, whether or not ISRE in lateral collisions may become dominant).

A second conclusion is that a correct treatment of the correlations between
internal variables and velocities at each point of space may be important;
this is illustrated by the figures contained in figure 2.\ In other words,
hydrodynamic equations are not always appropriate.\ What is remarkable is that
the effects of local field inhomogeneity and velocities combine non-linearly
to create this almost complete separation of the two internal states; the
effect is not a segregation of states but rather an internal conversion that
depends on the direction of the velocity.\ It is probably even more remarkable
that this effect was discovered experimentally, without any theoretical
prediction as a guide to the appropriate experimental conditions, and appeared
basically as a $100\%$ effect from the beginning.\ Clearly, ISRE is not a
small quantum correction to the effect of collisions, but may dominate the
entire dynamics of the quantum gas!

It would be interesting to extend the present work in a few directions, for
instance exploring the influence of a temperature gradient on the spin current
in hydrodynamical situations, as predicted in \cite{LL2}. For bosons, it would
also be important to understand better how spin waves evolve progressively
from the non-condensed regime, which we have studied in this article, to the
condensed low temperature regime; the theory recently developed in
\cite{Nikuni-Williams} seems to provide an appropriate tool for this purpose.

\bigskip

Acknowledgments: Le LKB est UMR 8552 du CNRS, de l'ENS et de l'Universit\'{e}
P. et M. Curie, Paris. Part of this work was made during a summer visit at the
ECT* center of theoretical physics near Trento (Italy), to which the authors
are very grateful.\ They also express their gratitude to J.\ Williams for many
stimulating discussions at ECT* and, later, by e-mail.

\begin{center}
\medskip\textbf{APPENDICES: INTERACTION\ TERM OF\ THE\ S-MATRIX\ ANSATZ}

\end{center}

\section*{Appendix I: spin-independent interactions}

In this appendix, we give the detailed calculation of the Wigner transform of
the collision term of the $S$ matrix Ansatz, equation (\ref{rhocoll}). We will
closely follow the notations of a similar calculation done in the Appendix of
Ref. \cite{Mullin-Laloe}, where the Wigner transform of the Yvon-Snider
Ansatz, equation (\ref{Yvon}) without internal levels and without exchange, is
obtained to first order in the gradient expansion. We define the on-shell $T$
matrix $\mathcal{T}$ by:
\begin{equation}
\hat{S}=\hat{1}-i2\pi\hat{\mathcal{T}} \label{ostm}%
\end{equation}
and rewrite the core of the Ansatz in equation (\ref{rhocoll}) as:
\begin{align}
\hat{S}\hat{\rho}_{1}(1)\hat{\rho}_{1}(2)\hat{S}^{\dagger}-\hat{\rho}%
_{1}(1)\hat{\rho}_{1}(2)  &  =-i2\pi\hat{\mathcal{T}}\hat{\rho}_{1}%
(1)\hat{\rho}_{1}(2)+\text{c.c.}\nonumber\\
&  +(2\pi)^{2}\hat{\mathcal{T}}\hat{\rho}_{1}(1)\hat{\rho}_{1}(2)\hat
{\mathcal{T}}^{\dagger}%
\end{align}
where c.c. is the complex conjugate of the preceding term. The first two terms
of the preceding equation are linear in the $T$ matrix, the last is quadratic.
The Wigner transform $\hat{F}_{T}$ of the linear in $T$ matrix operator
between curly brackets in (\ref{rhocoll}) is:
\begin{gather}
\hat{F}_{T}(\mathbf{R},\mathbf{r},\mathbf{P},\mathbf{p})=-i(2\pi)^{-5}%
\hbar^{-6}\int d^{3}K\int d^{3}\kappa~e^{i\mathbf{K}\cdot\mathbf{R}%
}e^{i\mathbf{\kappa}\cdot\mathbf{r}}\nonumber\\
\times\langle\mathbf{K}_{+},\mathbf{k}_{+}|\frac{\hat{1}+\epsilon\hat{P}%
_{ex.}}{\sqrt{2}}\hat{\mathcal{T}}\hat{\rho}_{1}(1)\hat{\rho}_{1}(2)\frac
{\hat{1}+\epsilon\hat{P}_{ex.}}{\sqrt{2}}|\mathbf{K}_{-},\mathbf{k}_{-}%
\rangle+\text{c.c.} \label{A2}%
\end{gather}
where:
\begin{equation}
\mathbf{K}_{\pm}=\frac{\mathbf{P}}{\hbar}\pm\frac{\mathbf{K}}{2}%
;\,\,\mathbf{k}_{\pm}=\frac{\mathbf{p}}{\hbar}\pm\frac{\mathbf{k}}{2}%
\end{equation}
With two closure relations for relative wavevectors $\mathbf{k}_{1}$ and
$\mathbf{k}_{2}$, the r.h.s. of (\ref{A2}) becomes equal to:%
\begin{gather}
\frac{1}{2i(2\pi)^{5}\hbar^{6}}\int d^{3}K\int d^{3}\kappa\int d^{3}k_{1}\int
d^{3}k_{2}~e^{i\mathbf{K}\cdot\mathbf{R}}e^{i\boldsymbol{\kappa}\cdot
\mathbf{r}}\nonumber\\
\times\left(  T(\mathbf{k}_{+},\mathbf{k}_{1})+\epsilon T(-\mathbf{k}%
_{+},\mathbf{k}_{1})\hat{P}_{S}\right)  \delta(E_{k_{+}}-E_{k_{1}})\nonumber\\
\times\langle\mathbf{K}_{+},\mathbf{k}_{1}|\hat{\rho}_{1}(1)\hat{\rho}%
_{1}(2)|\mathbf{K}_{-},\mathbf{k}_{2}\rangle\left(  \delta(\mathbf{k}%
_{2}-\mathbf{k}_{-})+\epsilon\delta(\mathbf{k}_{2}+\mathbf{k}_{-})\hat{P}%
_{S}\right)  +\text{c.c.} \label{A4}%
\end{gather}
where $\hat{P}_{S}$ is the exchange operator in spin space. Now using the
inverse Wigner transform formula, we can introduce into (\ref{A4}) the Wigner
transform $\hat{\rho}(\mathbf{r},\mathbf{p})$ of $\hat{\rho}_{1}$ and replace
the matrix element of the product of $\hat{\rho}$'s by:%
\begin{equation}%
\begin{array}
[c]{l}%
\hbar^{6}\int d^{3}R^{\prime}\int d^{3}r^{\prime}e^{-i\mathbf{K}%
\cdot\mathbf{R}^{\prime}}e^{i(\mathbf{k}_{2}-\mathbf{k}_{1})\cdot
\mathbf{r}^{\prime}}\\
\times\hat{\rho}\left(  \mathbf{R}^{\prime}+\frac{\mathbf{r}^{\prime}}%
{2},\frac{\mathbf{P}+\hbar\mathbf{k}_{1}+\hbar\mathbf{k}_{2}}{2}\right)
\hat{\rho}\left(  \mathbf{R}^{\prime}-\frac{\mathbf{r}^{\prime}}{2}%
,\frac{\mathbf{P}-\hbar\mathbf{k}_{1}-\hbar\mathbf{k}_{2}}{2}\right)
\end{array}
\end{equation}
When inserting the result into (\ref{A4}), a delta function appears:
\begin{equation}
\int d^{3}Ke^{i\mathbf{K}\cdot(\mathbf{R}-\mathbf{R}^{\prime})}=(2\pi
)^{3}\delta(\mathbf{R}-\mathbf{R}^{\prime})
\end{equation}
Tracing over the orbital and the spin space of particle $2$, we finally obtain
the following expression for the Wigner transform $\hat{I}_{T}(\mathbf{r}%
_{1},\mathbf{p}_{1})$ for the terms linear in $T$ in the Ansatz (\ref{rhocoll}%
):
\begin{align}
\hat{I}_{T}(\mathbf{r}_{1},\mathbf{p}_{1})  &  =\frac{1}{2i(2\pi)^{2}~\Delta
t}\int d^{3}q\int d^{3}r\int d^{3}\kappa\int d^{3}k_{1}^{\prime}\int
d^{3}k_{2}^{\prime}\int d^{3}r^{\prime}\nonumber\\
&  \times e^{i\boldsymbol{\kappa}\cdot\mathbf{r}}e^{i(\mathbf{k}_{2}^{\prime
}-\mathbf{k}_{1}^{\prime})\cdot\mathbf{r}^{\prime}}\delta(E_{k_{+}}%
-E_{k_{1}^{\prime}})\nonumber\\
&  \times\left[  \delta(\mathbf{k}_{2}^{\prime}-\mathbf{k}_{-})T(\mathbf{k}%
_{+},\mathbf{k}_{1}^{\prime})\hat{\rho}\left(  \mathbf{r}_{1}-\frac
{\mathbf{r}-\mathbf{r}^{\prime}}{2},\mathbf{p}_{1}^{\prime\prime}\right)
f\left(  \mathbf{r}_{1}-\frac{\mathbf{r}+\mathbf{r}^{\prime}}{2}%
,\mathbf{p}_{2}^{\prime\prime}\right)  \right. \nonumber\\
&  +\epsilon\delta(\mathbf{k}_{2}^{\prime}-\mathbf{k}_{-})T(-\mathbf{k}%
_{+},\mathbf{k}_{1}^{\prime})\hat{\rho}\left(  \mathbf{r}_{1}-\frac
{\mathbf{r}+\mathbf{r}^{\prime}}{2},\mathbf{p}_{2}^{\prime\prime}\right)
\hat{\rho}\left(  \mathbf{r}_{1}-\frac{\mathbf{r}-\mathbf{r}^{\prime}}%
{2},\mathbf{p}_{1}^{\prime\prime}\right) \nonumber\\
&  +\epsilon\delta(\mathbf{k}_{2}^{\prime}+\mathbf{k}_{-})T(\mathbf{k}%
_{+},\mathbf{k}_{1}^{\prime})\hat{\rho}\left(  \mathbf{r}_{1}-\frac
{\mathbf{r}-\mathbf{r}^{\prime}}{2},\mathbf{p}_{1}^{\prime\prime}\right)
\hat{\rho}\left(  \mathbf{r}_{1}-\frac{\mathbf{r}+\mathbf{r}^{\prime}}%
{2},\mathbf{p}_{2}^{\prime\prime}\right) \nonumber\\
&  +\left.  \delta(\mathbf{k}_{2}^{\prime}+\mathbf{k}_{-})T(-\mathbf{k}%
_{+},\mathbf{k}_{1}^{\prime})f\left(  \mathbf{r}_{1}-\frac{\mathbf{r}%
-\mathbf{r}^{\prime}}{2},\mathbf{p}_{1}^{\prime\prime}\right)  \hat{\rho
}\left(  \mathbf{r}_{1}-\frac{\mathbf{r}+\mathbf{r}^{\prime}}{2}%
,\mathbf{p}_{2}^{\prime\prime}\right)  \right] \nonumber\\
&  +\text{c.c.} \label{A9T}%
\end{align}
with notation (\ref{impulsions}) and with:
\begin{equation}%
\begin{array}
[c]{cc}%
\mathbf{r}=\mathbf{r}_{1}-\mathbf{r}_{2}\text{ ;} & \,\mathbf{k}_{\pm}%
=\frac{\mathbf{p}}{\hbar}\pm\frac{\boldsymbol{\kappa}}{2}\\
\mathbf{p}_{1}^{\prime\prime}=\mathbf{p}_{1}-\frac{\mathbf{q}}{2}+\hbar
\frac{\mathbf{k}_{1}^{\prime}+\mathbf{k}_{2}^{\prime}}{2}\text{ ;} &
\mathbf{p}_{2}^{\prime\prime}=\mathbf{p}_{1}-\frac{\mathbf{q}}{2}-\hbar
\frac{\mathbf{k}_{1}^{\prime}+\mathbf{k}_{2}^{\prime}}{2}%
\end{array}
\end{equation}
The distribution $f$ is the spin trace of $\hat{\rho}$, see equation
(\ref{def-f}).

The same kind of calculation can be done for the term quadratic in the $T$
matrix in (\ref{rhocoll}). It gives the Wigner transform $\hat{I}_{T^{2}%
}(\mathbf{r}_{1},\mathbf{p}_{1})$ as:
\begin{align}
\hat{I}_{T^{2}}(\mathbf{r}_{1},\mathbf{p}_{1})  &  =\frac{1}{4\pi\Delta t}\int
d^{3}q\int d^{3}r\int d^{3}\kappa\int d^{3}k_{1}^{\prime}\int d^{3}%
k_{2}^{\prime}\int d^{3}r^{\prime}\nonumber\\
&  \times e^{i\boldsymbol{\kappa}\cdot\mathbf{r}}~e^{i(\mathbf{k}_{2}^{\prime
}-\mathbf{k}_{1}^{\prime})\cdot\mathbf{r}^{\prime}}\delta(E_{k_{+}}%
-E_{k_{1}^{\prime}})\delta(E_{k_{-}}-E_{k_{2}^{\prime}})\nonumber\\
&  \times\left[  T(\mathbf{k}_{+},\mathbf{k}_{1}^{\prime})T(\mathbf{k}%
_{-},\mathbf{k}_{2}^{\prime})^{\ast}\hat{\rho}\left(  \mathbf{r}_{1}%
-\frac{\mathbf{r}-\mathbf{r}^{\prime}}{2},\mathbf{p}_{1}^{\prime\prime
}\right)  f\left(  \mathbf{r}_{1}-\frac{\mathbf{r}+\mathbf{r}^{\prime}}%
{2},\mathbf{p}_{2}^{\prime\prime}\right)  \right. \nonumber\\
&  +\epsilon T(-\mathbf{k}_{+},\mathbf{k}_{1}^{\prime})T(\mathbf{k}%
_{-},\mathbf{k}_{2}^{\prime})^{\ast}\hat{\rho}\left(  \mathbf{r}_{1}%
-\frac{\mathbf{r}-\mathbf{r}^{\prime}}{2},\mathbf{p}_{2}^{\prime\prime
}\right)  \hat{\rho}\left(  \mathbf{r}_{1}-\frac{\mathbf{r}+\mathbf{r}%
^{\prime}}{2},\mathbf{p}_{1}^{\prime\prime}\right) \nonumber\\
&  +\epsilon T(\mathbf{k}_{+},\mathbf{k}_{1}^{\prime})T(-\mathbf{k}%
_{-},\mathbf{k}_{2}^{\prime})^{\ast}\hat{\rho}\left(  \mathbf{r}_{1}%
-\frac{\mathbf{r}-\mathbf{r}^{\prime}}{2},\mathbf{p}_{1}^{\prime\prime
}\right)  \hat{\rho}\left(  \mathbf{r}_{1}-\frac{\mathbf{r}+\mathbf{r}%
^{\prime}}{2},\mathbf{p}_{2}^{\prime\prime}\right) \nonumber\\
&  +\left.  T(-\mathbf{k}_{+},\mathbf{k}_{1}^{\prime})T(-\mathbf{k}%
_{-},\mathbf{k}_{2}^{\prime})^{\ast}f\left(  \mathbf{r}_{1}-\frac
{\mathbf{r}-\mathbf{r}^{\prime}}{2},\mathbf{p}_{1}^{\prime\prime}\right)
\hat{\rho}\left(  \mathbf{r}_{1}-\frac{\mathbf{r}+\mathbf{r}^{\prime}}%
{2},\mathbf{p}_{2}^{\prime\prime}\right)  \right]  \label{A9T2}%
\end{align}
As in Ref. \cite{TS,Mullin-Laloe}, we now assume that $\hat{\rho}%
(\mathbf{r},\mathbf{p})$ varies slowly in space over microscopic distances and
expand the product of $\hat{\rho}$ and $f$ in (\ref{A9T}) and (\ref{A9T2})
according to
\begin{align}
&  \hat{\rho}(\mathbf{r}_{1},\mathbf{p}_{1}^{\prime\prime})f(\mathbf{r}%
_{1},\mathbf{p}_{2}^{\prime\prime})-\frac{\mathbf{r}}{2}\cdot\nabla
_{\mathbf{r}_{1}}\left[  \hat{\rho}(\mathbf{r}_{1},\mathbf{p}_{1}%
^{\prime\prime})f(\mathbf{r}_{1},\mathbf{p}_{2}^{\prime\prime})\right]
\nonumber\\
&  +\frac{\mathbf{r}^{\prime}}{2}\cdot\left[  \hat{\rho}(\mathbf{r}%
_{1},\mathbf{p}_{1}^{\prime\prime})\nabla_{\mathbf{r}_{1}}f(\mathbf{r}%
_{1},\mathbf{p}_{2}^{\prime\prime})-\left(  \nabla_{\mathbf{r}_{1}}\hat{\rho
}(\mathbf{r}_{1},\mathbf{p}_{1}^{\prime\prime})\right)  f(\mathbf{r}%
_{1},\mathbf{p}_{2}^{\prime\prime})+...\right]  \label{dvptgrad}%
\end{align}
The first term in (\ref{dvptgrad}) corresponds to the local term, the two
which follow to first order in gradients non-local terms. A very similar
equation exists for the product of $\hat{\rho}$'s in (\ref{A9T}) and
(\ref{A9T2}).

\paragraph{(1) Local term}

We now calculate the zeroth order in gradients. The following three integrals
occur:
\begin{align}
&  \int d^{3}r~e^{i\boldsymbol{\kappa}\cdot\mathbf{r}}=(2\pi)^{3}%
\delta(\boldsymbol{\kappa});\,\int d^{3}r^{\prime}~e^{i(\mathbf{k}_{2}%
^{\prime}-\mathbf{k}_{1}^{\prime})\cdot\mathbf{r}^{\prime}}=(2\pi)^{3}%
\delta(\mathbf{k}_{2}^{\prime}-\mathbf{k}_{1}^{\prime})\nonumber\\
&  \int d^{3}r^{\prime}~e^{-i(\mathbf{k}_{2}^{\prime}+\mathbf{k}_{1}^{\prime
})\cdot\mathbf{r}^{\prime}}=(2\pi)^{3}\delta(\mathbf{k}_{2}^{\prime
}+\mathbf{k}_{1}^{\prime})
\end{align}
The linear in $T$ matrix terms become:%
\begin{align}
\hat{I}_{T}(\mathbf{r}_{1},\mathbf{p}_{1})  &  =\frac{(2\pi)^{4}}{2i~\Delta
t}\int d^{3}q\int d^{3}k_{1}^{\prime}~\delta(E_{k}-E_{k_{1}^{\prime}%
})\nonumber\\
&  \times\Big[\delta(\mathbf{k}_{1}^{\prime}-\mathbf{k})T(\mathbf{k}%
,\mathbf{k}_{1}^{\prime})\hat{\rho}\left(  \mathbf{r}_{1},\mathbf{p}_{1}%
-\hbar\mathbf{k+}\hbar\mathbf{k}_{1}^{\prime}\right)  f\left(  \mathbf{r}%
_{1},\mathbf{p}_{1}-\hbar\mathbf{k-}\hbar\mathbf{k}_{1}^{\prime}\right)
\nonumber\\
&  +\epsilon\delta(\mathbf{k}_{1}^{\prime}-\mathbf{k})T(-\mathbf{k}%
,\mathbf{k}_{1}^{\prime})\hat{\rho}\left(  \mathbf{r}_{1},\mathbf{p}_{1}%
-\hbar\mathbf{k-}\hbar\mathbf{k}_{1}^{\prime}\right)  \hat{\rho}\left(
\mathbf{r}_{1},\mathbf{p}_{1}-\hbar\mathbf{k+}\hbar\mathbf{k}_{1}^{\prime
}\right) \nonumber\\
&  +\epsilon\delta(\mathbf{k}_{1}^{\prime}+\mathbf{k})T(\mathbf{k}%
,\mathbf{k}_{1}^{\prime})\hat{\rho}\left(  \mathbf{r}_{1},\mathbf{p}_{1}%
-\hbar\mathbf{k+}\hbar\mathbf{k}_{1}^{\prime}\right)  \hat{\rho}\left(
\mathbf{r}_{1},\mathbf{p}_{1}-\hbar\mathbf{k-}\hbar\mathbf{k}_{1}^{\prime
}\right) \nonumber\\
&  +\delta(\mathbf{k}_{1}^{\prime}+\mathbf{k})T(-\mathbf{k},\mathbf{k}%
_{1}^{\prime})f\left(  \mathbf{r}_{1},\mathbf{p}_{1}-\hbar\mathbf{k+}%
\hbar\mathbf{k}_{1}^{\prime}\right)  \hat{\rho}\left(  \mathbf{r}%
_{1},\mathbf{p}_{1}-\hbar\mathbf{k-}\hbar\mathbf{k}_{1}^{\prime}\right)
\Big ]\nonumber\\
&  +\text{c.c.} \label{A17T}%
\end{align}
In the preceding equation, we note the appearance of the square of delta
functions of the energy $\delta(E_{k})\delta(\mathbf{k})\propto\left[
\delta(E_{k})\right]  ^{2}$. They are handled by the following well-known
simplification from scattering theory:
\begin{equation}
\left[  \delta(E_{k})\right]  ^{2}=\frac{\Delta t}{2\pi\hbar}\delta(E_{k})
\label{delta2}%
\end{equation}
where $\Delta t$ is a time larger than the duration of a collision (see Ref.
\cite{Roman}, for example), which simplifies with the one introduced in the
$S$ matrix Ansatz (\ref{rhocoll}). The same manipulations can be done with the
quadratic in $T$ matrix term, they lead to:%
\begin{align}
\hat{I}_{T^{2}}(\mathbf{r}_{1},\mathbf{p}_{1})  &  =\frac{(2\pi)^{4}}{2\hbar
}\int d^{3}q\int d^{3}k^{\prime}~\delta(E_{k}-E_{k^{\prime}})\nonumber\\
&  \times\Big[T(\mathbf{k},\mathbf{k}^{\prime})T(\mathbf{k},\mathbf{k}%
^{\prime})^{\ast}\hat{\rho}\left(  \mathbf{r}_{1},\mathbf{p}_{1}%
-\hbar\mathbf{k}+\hbar\mathbf{k}^{\prime}\right)  f\left(  \mathbf{r}%
_{1},\mathbf{p}_{1}-\hbar\mathbf{k}-\hbar\mathbf{k}^{\prime}\right)
\nonumber\\
&  +\epsilon T(-\mathbf{k},\mathbf{k}^{\prime})T(\mathbf{k},\mathbf{k}%
^{\prime})^{\ast}\hat{\rho}\left(  \mathbf{r}_{1},\mathbf{p}_{1}%
-\hbar\mathbf{k}-\hbar\mathbf{k}^{\prime}\right)  \hat{\rho}\left(
\mathbf{r}_{1},\mathbf{p}_{1}-\hbar\mathbf{k}+\hbar\mathbf{k}^{\prime}\right)
\nonumber\\
&  +\epsilon T(\mathbf{k},\mathbf{k}^{\prime})T(-\mathbf{k},\mathbf{k}%
^{\prime})^{\ast}\hat{\rho}\left(  \mathbf{r}_{1},\mathbf{p}_{1}%
-\hbar\mathbf{k}+\hbar\mathbf{k}^{\prime}\right)  \hat{\rho}\left(
\mathbf{r}_{1},\mathbf{p}_{1}-\hbar\mathbf{k}-\hbar\mathbf{k}^{\prime}\right)
\nonumber\\
&  +T(-\mathbf{k},\mathbf{k}^{\prime})T(-\mathbf{k},\mathbf{k}^{\prime}%
)^{\ast}f\left(  \mathbf{r}_{1},\mathbf{p}_{1}-\hbar\mathbf{k}+\hbar
\mathbf{k}^{\prime}\right)  \hat{\rho}\left(  \mathbf{r}_{1},\mathbf{p}%
_{1}-\hbar\mathbf{k}-\hbar\mathbf{k}^{\prime}\right)  \Big] \label{A17T2}%
\end{align}

We will now use different properties of the $T$ matrix. In addition to the
differential $\sigma_{k}(\hat{k},\hat{k}^{\prime})$ and total $\sigma_{T}(k)$
cross sections defined in (\ref{sigmas}), we introduce, following Ref.
\cite{LL1}, the following cross sections:
\begin{align}
T(-\mathbf{k},\mathbf{k})  &  =\frac{\hbar^{2}k}{i8\pi^{3}m}\left(
\sigma_{fwd.}^{ex.}(k)-i\tau_{fwd.}^{ex.}(k)\right) \nonumber\\
T(-\mathbf{k}^{\prime},\mathbf{k})T(\mathbf{k}^{\prime},\mathbf{k})^{\ast}  &
=\frac{\hbar^{4}}{4\pi^{4}m^{2}}\left(  \sigma_{k}^{ex.}(\hat{k},\hat
{k}^{\prime})-i\tau_{k}^{ex.}(\hat{k},\hat{k}^{\prime})\right)
\label{sigmas2}%
\end{align}
Unitarity of the $S$ matrix implies the optical theorem:
\begin{equation}
\frac{T(\mathbf{k},\mathbf{k})}{2i}+c.c.=\operatorname{Im}T(\mathbf{k}%
,\mathbf{k})=-\frac{\hbar^{2}k}{(2\pi)^{3}m}\sigma_{T}(k) \label{opt}%
\end{equation}
and the rotational invariance of the interaction Hamiltonian can be used to
show that:
\begin{equation}
T(-\mathbf{k},-\mathbf{k}^{\prime})=T(\mathbf{k},\mathbf{k}^{\prime})
\label{rot}%
\end{equation}
Performing the integral over the length of wavevector $\mathbf{k}^{\prime}$ in
(\ref{A17T}) and (\ref{A17T2}) and using the previous properties of the $T$
matrix, we obtain the Wigner transform to zero order of the $S$ matrix Ansatz
$\hat{I}_{W}=\hat{I}_{T}+\hat{I}_{T^{2}}$:
\begin{align}
&  \hat{I}_{W}(\mathbf{r}_{1},\mathbf{p}_{1})=-\int d^{3}q\,\frac{q}%
{m}\Bigg\{\bigg[\sigma_{T}(k)\hat{\rho}(\mathbf{r}_{1},\mathbf{p}%
_{1})f(\mathbf{r}_{1},\mathbf{p}_{2})\nonumber\\
&  +\frac{\epsilon}{2}\left(  \sigma_{fwd.}^{ex.}(k)\left[  \hat{\rho
}(\mathbf{r}_{1},\mathbf{p}_{1}),\hat{\rho}(\mathbf{r}_{1},\mathbf{p}%
_{2})\right]  _{+}+i\tau_{fwd.}^{ex.}(k)\left[  \hat{\rho}(\mathbf{r}%
_{1},\mathbf{p}_{1}),\hat{\rho}(\mathbf{r}_{1},\mathbf{p}_{2})\right]
_{-}\right)  \bigg]\nonumber\\
&  -\int d^{2}\hat{k}^{\prime}\bigg[\sigma_{k}(\hat{k},\hat{k}^{\prime}%
)\hat{\rho}(\mathbf{r}_{1},\mathbf{p}_{1}^{\prime})f(\mathbf{r}_{1}%
,\mathbf{p}_{2}^{\prime})+\frac{\epsilon}{2}\Big(\sigma_{k}^{ex.}(\hat{k}%
,\hat{k}^{\prime})\left[  \hat{\rho}(\mathbf{r}_{1},\mathbf{p}_{1}^{\prime
}),\hat{\rho}(\mathbf{r}_{1},\mathbf{p}_{2}^{\prime})\right]  _{+}\nonumber\\
&  +i\tau_{k}^{ex.}(\hat{k},\hat{k}^{\prime})\left[  \hat{\rho}(\mathbf{r}%
_{1},\mathbf{p}_{1}^{\prime}),\hat{\rho}(\mathbf{r}_{1},\mathbf{p}_{2}%
^{\prime})\right]  _{-}\Big)\bigg]\Bigg\} \label{ILL}%
\end{align}
with the notations (\ref{impulsions}). This local term is the LL
\textquotedblleft collision integral\textquotedblright\ \cite{LL1}. It
contains terms linear in $T$ matrix, usually written on the l.h.s. of the
kinetic equation, as well as terms quadratic in $T$ matrix.

The low-energy limit of the cross sections is obtained from the low-energy
expression of the $T$ matrix (\ref{1}):%
\[%
\begin{array}
[c]{cc}%
\sigma_{T}(k)\sim4\pi a^{2} & \sigma_{k}(\hat{k},\hat{k}^{\prime})\sim a^{2}\\
\sigma_{fwd.}^{ex.}(k)\sim4\pi a^{2} & \sigma_{k}^{ex.}(\hat{k},\hat
{k}^{\prime})\sim a^{2}\\
\tau_{fwd.}^{ex.}(k)\sim-4\pi a/k & \tau_{k}^{ex.}(\hat{k},\hat{k}^{\prime
})\rightarrow0
\end{array}
\]
when $k\rightarrow0$, see Ref. \cite{LL1}. When introduced into (\ref{ILL}),
the collision integral of equation (\ref{Icoll}) is obtained, as well as the
spin mean field contained in the commutator in equation (\ref{kinetic0}).
Other mean field terms (contained in the anticommutator) appear to first order
of the gradient expansion.

\paragraph{(2) First order terms}

We retain only first order terms that are in addition linear in $T$ matrix.

\subparagraph{(2-a) $\mathbf{r}$ gradients}

The first-order terms introduce the gradient of a delta function:%
\[
\int d^{3}r~\mathbf{r}\text{ }e^{i\boldsymbol{\kappa}\cdot\mathbf{r}}%
=-i(2\pi)^{3}\mathbf{\nabla}_{\boldsymbol{\kappa}}\delta(\boldsymbol{\kappa})
\]
which implies taking the derivative with respect to $\boldsymbol{\kappa}$ of
the function under the integral. We obtain:%
\begin{align}
\frac{(2\pi)^{3}}{2}  &  \int d^{3}q\bigg\{\operatorname{Re}T(\mathbf{k}%
,\mathbf{k})\text{ }\mathbf{\nabla}_{\mathbf{p}}\cdot\mathbf{\nabla
}_{\mathbf{r}}\left[  \hat{\rho}(\mathbf{r}_{1},\mathbf{p}_{1})f(\mathbf{r}%
_{1},\mathbf{p}_{2})\right] \nonumber\\
&  -\mathbf{\nabla}_{\mathbf{p}}\left[  \operatorname{Re}T(\mathbf{k}%
,\mathbf{k})\right]  \text{ }\cdot\mathbf{\nabla}_{\mathbf{r}}\left[
\hat{\rho}(\mathbf{r}_{1},\mathbf{p}_{1})f(\mathbf{r}_{1},\mathbf{p}%
_{2})\right] \nonumber\\
&  +\frac{\epsilon}{2}T(\mathbf{k},\mathbf{k})\text{ }\mathbf{\nabla
}_{\mathbf{p}}\cdot\mathbf{\nabla}_{\mathbf{r}}\left[  \hat{\rho}%
(\mathbf{r}_{1},\mathbf{p}_{1})\hat{\rho}(\mathbf{r}_{1},\mathbf{p}%
_{2})\right] \nonumber\\
&  -\frac{\epsilon}{2}\mathbf{\nabla}_{\mathbf{p}}\left[  T(\mathbf{k}%
,\mathbf{k})\right]  \cdot\mathbf{\nabla}_{\mathbf{r}}\left[  \hat{\rho
}(\mathbf{r}_{1},\mathbf{p}_{1})\hat{\rho}(\mathbf{r}_{1},\mathbf{p}%
_{2})\right] \nonumber\\
&  +\frac{\epsilon}{2}T(\mathbf{k},\mathbf{k})^{\ast}\text{ }\mathbf{\nabla
}_{\mathbf{p}}\cdot\mathbf{\nabla}_{\mathbf{r}}\left[  \hat{\rho}%
(\mathbf{r}_{1},\mathbf{p}_{2})\hat{\rho}(\mathbf{r}_{1},\mathbf{p}%
_{1})\right] \nonumber\\
&  -\frac{\epsilon}{2}\mathbf{\nabla}_{\mathbf{p}}\left[  T(\mathbf{k}%
,\mathbf{k})^{\ast}\right]  \cdot\mathbf{\nabla}_{\mathbf{r}}\left[  \hat
{\rho}(\mathbf{r}_{1},\mathbf{p}_{2})\hat{\rho}(\mathbf{r}_{1},\mathbf{p}%
_{1})\right]  \bigg\} \label{A20}%
\end{align}

\subparagraph{(2-b) $\mathbf{r}^{\prime}$ gradients}

The following integrals occur:
\begin{align}
&  \int d^{3}r^{\prime}~e^{i(\mathbf{k}_{2}^{\prime}-\mathbf{k}_{1}^{\prime
})\cdot\mathbf{r}^{\prime}}\mathbf{r}^{\prime}=-i(2\pi)^{3}\mathbf{\nabla
}_{\mathbf{k}_{1}^{\prime}}\delta(\mathbf{k}_{1}^{\prime}-\mathbf{k}%
_{2}^{\prime})\nonumber\\
&  \int d^{3}r^{\prime}~e^{-i(\mathbf{k}_{2}^{\prime}+\mathbf{k}_{1}^{\prime
})\cdot\mathbf{r}^{\prime}}\mathbf{r}^{\prime}=-i(2\pi)^{3}\mathbf{\nabla
}_{\mathbf{k}_{1}^{\prime}}\delta(\mathbf{k}_{2}^{\prime}+\mathbf{k}%
_{1}^{\prime})
\end{align}
so that we obtain:
\begin{align}
\frac{(2\pi)^{3}}{2}  &  \int d^{3}q~\mathbf{\nabla}_{\mathbf{p}}%
\cdot\bigg \{\operatorname{Re}T(\mathbf{k},\mathbf{k})\text{ }\left[
\hat{\rho}(\mathbf{r}_{1},\mathbf{p}_{1})\mathbf{\nabla}_{\mathbf{r}%
}f(\mathbf{r}_{1},\mathbf{p}_{2})-f(\mathbf{r}_{1},\mathbf{p}_{2}%
)\mathbf{\nabla}_{\mathbf{r}}\hat{\rho}(\mathbf{r}_{1},\mathbf{p}_{1})\right]
\nonumber\\
&  +\frac{\epsilon}{2}T(\mathbf{k},-\mathbf{k})\text{ }\left[  \hat{\rho
}(\mathbf{r}_{1},\mathbf{p}_{1})\mathbf{\nabla}_{\mathbf{r}}\hat{\rho
}(\mathbf{r}_{1},\mathbf{p}_{2})-\hat{\rho}(\mathbf{r}_{1},\mathbf{p}%
_{2})\mathbf{\nabla}_{\mathbf{r}}\hat{\rho}(\mathbf{r}_{1},\mathbf{p}%
_{1})\right] \nonumber\\
&  +\frac{\epsilon}{2}T(\mathbf{k},-\mathbf{k})^{\ast}\text{ }\left[
\mathbf{\nabla}_{\mathbf{r}}\hat{\rho}(\mathbf{r}_{1},\mathbf{p}_{2})\text{
}\hat{\rho}(\mathbf{r}_{1},\mathbf{p}_{1})-\mathbf{\nabla}_{\mathbf{r}}%
\hat{\rho}(\mathbf{r}_{1},\mathbf{p}_{1})\text{ }\hat{\rho}(\mathbf{r}%
_{1},\mathbf{p}_{2})\right]  \bigg\} \label{A21}%
\end{align}

The expression of the $T$ matrix to lowest order in the scattering length (see
equation (\ref{1}))
\[
(2\pi)^{3}T(\mathbf{k},\mathbf{k})\simeq g
\]
is independent of $\mathbf{k}$. Using the preceding expression in (\ref{A20})
and (\ref{A21}), we obtain the following first order term:
\[
\frac{1}{2}\left[  \nabla_{\mathbf{p}}\hat{\rho}(\mathbf{r}_{1},\mathbf{p}%
_{1}),\cdot\nabla_{\mathbf{r}}\big(gn(\mathbf{r}_{1})\hat{1}+\epsilon
\,g\hat{n}(\mathbf{r}_{1})\big)\right]  _{+}
\]
This is the mean field term appearing in the anticommutator in the kinetic
equation (\ref{kinetic0}).

\section*{Appendix II: spin-dependent interactions (forward scattering)}

In this appendix, we show how to obtain the full mean field in the case of
spin-dependent interactions, using the $S$-matrix Ansatz. We will limit
ourselves to collisions at low-energy, using the $T$ matrix only at lowest
order in the scattering lengths (equation (\ref{Tspin})):
\begin{equation}
\langle\alpha~;\beta|~\hat{T}_{k}~|\gamma~;\delta\rangle=\delta_{\alpha
,\gamma}\delta_{\beta,\delta}\frac{g_{\alpha\beta}^{(d)}}{(2\pi)^{3}%
}+(1-\delta_{\alpha,\beta})\delta_{\alpha,\delta}\delta_{\beta,\gamma}%
\frac{g_{\alpha\beta}^{(t)}}{(2\pi)^{3}} \label{Tind}%
\end{equation}
where $g_{\alpha\alpha}^{(d)}\equiv g_{\alpha\alpha}$. In this limit the $T$
matrix elements are real.

As we are interested in mean field terms, we only keep terms linear in the $T$
matrix in the r.h.s. of the $S$ matrix Ansatz, equation (\ref{rhocoll}):
\begin{equation}
\frac{2\pi}{i\Delta t}Tr_{2}\left\{  \frac{\hat{1}+\epsilon\hat{P}_{ex.}%
}{\sqrt{2}}\left[  \hat{\mathcal{T}}~\hat{\rho}_{1}(1)~\hat{\rho}_{1}%
(2)-\hat{\rho}_{1}(1)~\hat{\rho}_{1}(2)~\hat{\mathcal{T}}\right]  \frac
{\hat{1}+\epsilon\hat{P}_{ex.}}{\sqrt{2}}\right\}
\end{equation}
where we introduced the $\mathcal{T}$ matrix (the on-shell $T$ matrix), whose
definition is given in Appendix I, equation (\ref{ostm}). Using the properties
of the trace on particle $2$ and the fact that the exchange operator commutes
with the $\mathcal{T}$ matrix, we can rewrite the preceding equation in the
form of a commutator:
\begin{equation}
\frac{1}{i\hbar}\bigg[\frac{2\pi\hbar}{\Delta t}Tr_{2}\left\{  (\hat
{1}+\epsilon\hat{P}_{ex.})\hat{\mathcal{T}}~\hat{\rho}_{1}(2)\right\}
,~\hat{\rho}_{1}(1)\bigg]_{-} \label{commutator}%
\end{equation}
It seems therefore natural that the first operator in the commutator should
play the role of an effective single-particle Hamiltonian.

We now consider a general matrix element of the preceding commutator. To
simplify the notation, we only write the first term of the commutator:
\begin{equation}
\langle1:\mathbf{p}_{1},\alpha|\frac{2\pi\hbar}{\Delta t}Tr_{2}\left\{
(\hat{1}+\epsilon\hat{P}_{ex.})\hat{\mathcal{T}}~\hat{\rho}_{1}(2)\right\}
\hat{\rho}_{1}(1)|1:\mathbf{p}_{1}^{\prime},\alpha^{\prime}\rangle
\label{matelem}%
\end{equation}
We calculate this matrix element by introducing three closure relations of the
form:
\begin{equation}
\hat{1}=\int d^{3}p\sum_{\beta}|\mathbf{p},\beta\rangle\langle\mathbf{p}%
,\beta|
\end{equation}
and by using the expression (\ref{Tind}) of the $T$ matrix. We obtain:
\begin{align}
\frac{2\pi\hbar}{(2\pi)^{3}\Delta t}  &  \times\int d^{3}p_{2}\int d^{3}%
p_{3}~\delta(E_{(\mathbf{p}_{1}-\mathbf{p}_{2})/2}-E_{(\mathbf{p}%
_{3}-\mathbf{p}_{4})/2})\nonumber\\
&  \times\sum_{\beta}\bigg\{g_{\alpha\beta}^{(d)}\langle\mathbf{p}_{3}%
,\alpha|\hat{\rho}_{1}|\mathbf{p}_{1}^{\prime},\alpha^{\prime}\rangle
\langle\mathbf{p}_{4},\beta|\hat{\rho}_{1}|\mathbf{p}_{2},\beta\rangle
\nonumber\\
&  +(1-\delta_{\alpha,\beta})g_{\alpha\beta}^{(t)}\langle\mathbf{p}_{3}%
,\beta|\hat{\rho}_{1}|\mathbf{p}_{1}^{\prime},\alpha^{\prime}\rangle
\langle\mathbf{p}_{4},\alpha|\hat{\rho}_{1}|\mathbf{p}_{2},\beta
\rangle\nonumber\\
&  +\epsilon g_{\alpha\beta}^{(d)}\langle\mathbf{p}_{3},\beta|\hat{\rho}%
_{1}|\mathbf{p}_{1}^{\prime},\alpha^{\prime}\rangle\langle\mathbf{p}%
_{4},\alpha|\hat{\rho}_{1}|\mathbf{p}_{2},\beta\rangle\nonumber\\
&  +\epsilon(1-\delta_{\alpha,\beta})g_{\alpha\beta}^{(t)}\langle
\mathbf{p}_{3},\alpha|\hat{\rho}_{1}|\mathbf{p}_{1}^{\prime},\alpha^{\prime
}\rangle\langle\mathbf{p}_{4},\beta|\hat{\rho}_{1}|\mathbf{p}_{2},\beta
\rangle\bigg\} \label{tgre}%
\end{align}
where $\mathbf{p}_{4}=\mathbf{p}_{1}+\mathbf{p}_{2}-\mathbf{p}_{3}$. By
defining the three following coupling constants:%
\begin{equation}
g_{\alpha\alpha}=g_{\alpha\alpha}^{(d)}~~~\ \ ~~~;~~~~~~~g_{\alpha\beta
}=g_{\alpha\beta}^{(d)}+\epsilon g_{\alpha,\beta}^{(t)}~~~~~\text{if }%
\alpha\neq\beta
\end{equation}
we can rewrite equation (\ref{tgre}) as:
\begin{align}
\frac{2\pi\hbar}{\Delta t}  &  \times\int d^{3}p_{2}\int d^{3}p_{3}%
~\delta(E_{(\mathbf{p}_{1}-\mathbf{p}_{2})/2}-E_{(\mathbf{p}_{3}%
-\mathbf{p}_{4})/2})\nonumber\\
&  \times\sum_{\beta}\frac{g_{\alpha\beta}}{(2\pi)^{3}}\bigg\{\langle
\mathbf{p}_{4},\beta|\hat{\rho}_{1}|\mathbf{p}_{2},\beta\rangle\langle
\mathbf{p}_{3},\alpha|\hat{\rho}_{1}|\mathbf{p}_{1}^{\prime},\alpha^{\prime
}\rangle\nonumber\\
&  +\epsilon\langle\mathbf{p}_{4},\alpha|\hat{\rho}_{1}|\mathbf{p}_{2}%
,\beta\rangle\langle\mathbf{p}_{3},\beta|\hat{\rho}_{1}|\mathbf{p}_{1}%
^{\prime},\alpha^{\prime}\rangle\bigg\}
\end{align}
We therefore see that, for both statistics (bosonic or fermionic), only three
coupling constants ($g_{11}$, $g_{22}$ and $g_{12}$) are involved.

Now, taking the Wigner transform of the operator (\ref{commutator}) and
expanding in gradients (see Appendix I) introduces another delta function of
energy conservation, which is present at each order in gradients. This allows
to use the \textquotedblleft$\delta(E)^{2}$\textquotedblright\ simplification
(equation (\ref{delta2}) in Appendix I) to make the $\Delta t$ disappear. The
net result is that:
\begin{equation}
\frac{2\pi\hbar}{\Delta t}\delta(E_{(\mathbf{p}_{1}-\mathbf{p}_{2}%
)/2}-E_{(\mathbf{p}_{3}-\mathbf{p}_{4})/2})
\end{equation}
gets replaced by $1$. Formally the matrix element of equation (\ref{matelem})
becomes:
\begin{align}
\int d^{3}p_{2}\int d^{3}p_{3}\sum_{\beta}\frac{g_{\alpha\beta}}{(2\pi)^{3}}
&  \times\bigg\{\langle\mathbf{p}_{4},\beta|\hat{\rho}_{1}|\mathbf{p}%
_{2},\beta\rangle\langle\mathbf{p}_{3},\alpha|\hat{\rho}_{1}|\mathbf{p}%
_{1}^{\prime},\alpha^{\prime}\rangle\nonumber\\
&  +\epsilon\langle\mathbf{p}_{4},\alpha|\hat{\rho}_{1}|\mathbf{p}_{2}%
,\beta\rangle\langle\mathbf{p}_{3},\beta|\hat{\rho}_{1}|\mathbf{p}_{1}%
^{\prime},\alpha^{\prime}\rangle\bigg\}
\end{align}
This result can be rewritten:
\begin{equation}
\langle1:\mathbf{p}_{1},\alpha|\hat{V}^{mf}(1)\hat{\rho}_{1}(1)|1:\mathbf{p}%
_{1}^{\prime},\alpha^{\prime}\rangle
\end{equation}
where:
\begin{equation}
V_{\alpha\beta}^{mf}(\mathbf{r})=\delta_{\alpha\beta}\sum_{\gamma
=1,2}g_{\alpha\gamma}\langle\mathbf{r},\gamma|\hat{\rho}_{1}|\mathbf{r}%
,\gamma\rangle+\epsilon g_{\alpha\beta}\langle\mathbf{r},\alpha|\hat{\rho}%
_{1}|\mathbf{r},\beta\rangle
\end{equation}
is the mean field potential. Therefore, equation (\ref{commutator}) can be
written formally:
\begin{equation}
\frac{1}{i\hbar}\Big[\hat{V}^{mf}(1),~\hat{\rho}_{1}(1)\Big]_{-}%
\end{equation}
This equation is valid provided it is used to compute the Wigner transform and
do a subsequent gradient expansion.

In the body of the article, we define the effective single-particle
Hamiltonian $\hat{U}$ which is the sum of the external potential $\hat
{V}^{ext}$ and of the mean field potential $\hat{V}^{mf}$. The operatorial
kinetic equation is now:
\begin{equation}
\frac{d\hat{\rho}_{1}}{dt}+\left(  i\hbar\right)  ^{-1}~\left[  \hat{\rho}%
_{1},\hat{H}_{1}+\hat{V}^{mf}\right]  _{-}=0
\end{equation}
The kinetic equation is then derived following the procedure of Appendix I
(Wigner transform and gradient expansion up to first order); the result is:
\begin{align}
\partial_{t}\hat{\rho}(\mathbf{r},\mathbf{p})  &  +\frac{\mathbf{p}}{m}%
\cdot\nabla_{\mathbf{r}}\hat{\rho}(\mathbf{r},\mathbf{p})+\frac{1}{i\hbar
}[\hat{\rho}(\mathbf{r},\mathbf{p}),\widehat{U}(\mathbf{r})]_{-}\nonumber\\
&  -\frac{1}{2}[\nabla_{\mathbf{p}}\hat{\rho}(\mathbf{r},\mathbf{p}%
),\cdot\nabla_{\mathbf{r}}\widehat{U}(\mathbf{r})]_{+}=0
\end{align}
which is equal to the l.h.s. of (\ref{kinetic}).

\end{document}